\PassOptionsToPackage{unicode}{hyperref}
\PassOptionsToPackage{hyphens}{url}
\documentclass[
  american,
  11pt,
]{article}
\usepackage{xcolor}
\usepackage[margin=25mm]{geometry}
\usepackage{amsmath,amssymb}
\usepackage{iftex}
\ifPDFTeX
  \usepackage[T1]{fontenc}
  \usepackage[utf8]{inputenc}
  \usepackage{textcomp} 
\else 
  \usepackage{unicode-math} 
  \defaultfontfeatures{Scale=MatchLowercase}
  \defaultfontfeatures[\rmfamily]{Ligatures=TeX,Scale=1}
\fi
\usepackage{lmodern}

\IfFileExists{upquote.sty}{\usepackage{upquote}}{}
\IfFileExists{microtype.sty}{
  \usepackage[]{microtype}
  \UseMicrotypeSet[protrusion]{basicmath} 
}{}
\makeatletter
\@ifundefined{KOMAClassName}{
  \IfFileExists{parskip.sty}{%
    \usepackage{parskip}
  }{
    \setlength{\parindent}{0pt}
    \setlength{\parskip}{6pt plus 2pt minus 1pt}}
}{
  \KOMAoptions{parskip=half}}
\makeatother
\usepackage{longtable,booktabs,array}
\usepackage{calc} 
\usepackage{etoolbox}
\makeatletter
\patchcmd\longtable{\par}{\if@noskipsec\mbox{}\fi\par}{}{}
\makeatother
\IfFileExists{footnotehyper.sty}{\usepackage{footnotehyper}}{\usepackage{footnote}}
\makesavenoteenv{longtable}
\usepackage{graphicx}
\makeatletter
\newsavebox\pandoc@box
\newcommand*\pandocbounded[1]{
  \sbox\pandoc@box{#1}%
  \Gscale@div\@tempa{\textheight}{\dimexpr\ht\pandoc@box+\dp\pandoc@box\relax}%
  \Gscale@div\@tempb{\linewidth}{\wd\pandoc@box}%
  \ifdim\@tempb\p@<\@tempa\p@\let\@tempa\@tempb\fi
  \ifdim\@tempa\p@<\p@\scalebox{\@tempa}{\usebox\pandoc@box}%
  \else\usebox{\pandoc@box}%
  \fi%
}
\def\fps@figure{htbp}
\makeatother
\ifLuaTeX
\usepackage[bidi=basic,shorthands=off]{babel}
\else
\usepackage[bidi=default,shorthands=off]{babel}
\fi

\ifLuaTeX
  \usepackage{selnolig} 
\fi
\providecommand{\tightlist}{%
  \setlength{\itemsep}{0pt}\setlength{\parskip}{0pt}}
\usepackage{lineno}
\usepackage{setspace}
\usepackage{float}
\usepackage{xurl}
\usepackage{algorithm}
\usepackage{algpseudocode}
\algrenewcommand\algorithmicrequire{\textbf{Input:}}
\algrenewcommand\algorithmicensure{\textbf{Output:}}
\usepackage{bookmark}
\IfFileExists{xurl.sty}{\usepackage{xurl}}{} 
\hypersetup{
  pdftitle={Localizing and quantifying cerebral arterial stenosis from sparse velocity waveforms using a conditional neural operator},
  pdflang={en-US},
  hidelinks,
  pdfcreator={LaTeX via pandoc}}

\usepackage{authblk}

\title{Localizing and quantifying cerebral arterial stenosis from sparse velocity waveforms using a conditional neural operator}

\author[1]{Tong Zhu}
\author[2]{Yi Zhu}
\author[3,*]{Jiawei Liu}
\author[2,*]{Pipi Hu}

\affil[1]{School of Materials Science and Engineering,
Hefei University of Technology, Hefei, China}

\affil[2]{Beijing Institute of Mathematical Sciences and
Applications, Beijing 101408, China}

\affil[3]{Advanced Institute for Materials Research,
Tohoku University, Sendai, Japan}

\affil[*]{Corresponding authors:
Jiawei Liu (\texttt{liu.jiawei.e2@tohoku.ac.jp});
Pipi Hu (\texttt{hpp@bimsa.cn})}

\date{}

\begin{document}
\maketitle

\textbf{Short title:} Cerebral stenosis inversion from sparse velocity
waveforms

\section{Abstract}\label{abstract}

Cerebral arterial stenosis alters blood flow throughout the vascular
network, complicating the estimation of its location and severity from
sparse velocity observations. We developed a conditional neural operator
that predicts cerebral hemodynamics for candidate stenoses and enables
inverse estimation through waveform comparison. Training data were
generated using a sequential 0D--1D model that simulates pressure,
cross-sectional area, and velocity under varying anatomical,
physiological, and stenosis conditions. The operator encodes a single
stenosis within vessel tokens and uses global attention to represent its
effects across the network. Severity-gated coupling connects these
representations to a shared field decoder, while paired healthy
simulations constrain baseline predictions. After training, model weights
remain fixed, and stenosis location and severity are estimated by
minimizing the mismatch between predicted and observed midpoint velocity
waveforms. Network-averaged relative validation errors were 0.14\% for
pressure, 0.59\% for area, and 2.79\% for velocity. For an eight-vessel observation configuration, localization accuracy across 20 candidate
vessels was 99.4\% without added noise and 96.0\% at the 30\% noise
setting. Comparisons between configurations with equal numbers of
observed vessels showed that measurement placement influenced recovery.
Larger field reconstruction errors and greater localization ambiguities
were observed in selected communicating and distal arteries. These
results demonstrate the feasibility of surrogate-based stenosis
estimation in simulated cerebral networks and highlight the importance
of observation placement. Evaluation was limited to unseen stenosis
scenarios within sampled anatomical and physiological conditions.
Generalization to new individuals and clinical measurements requires
further investigation.

\section{Author Summary}\label{author-summary}

A narrowing in one artery can change blood flow in several connected
vessels. This creates both an opportunity and a difficulty: measurements
away from the narrowing may contain information about it, but different
narrowings can produce similar signals. We developed a computational
approach that learns how cerebral blood flow changes with the location
and severity of an arterial narrowing. We first generated examples with
a blood-flow simulator and trained a neural network to predict pressure,
vessel area, and velocity. We then used the trained network to compare
possible narrowings against velocity measurements available at only a
few vessels. In our simulated experiments, the method identified most
narrowing locations, including when noise was added to the observations.
Both the location and number of measurements mattered, and some
vessels remained harder to distinguish than others. Our study connects a
computational method for estimating disease parameters with an analysis
of where sparse measurements are informative. The results provide a
basis for further testing with new anatomies and experimental
measurements; they do not yet establish diagnostic performance in
patients.

\section{1. Introduction}\label{introduction}

Cerebral arterial stenosis is a localized reduction in lumen diameter whose hemodynamic consequences depend on the surrounding vascular network. The communicating arteries
of the Circle of Willis can redistribute flow when a major feeding artery is narrowed; consequently, a measured waveform reflects both local vascular properties and network interactions {[}1,2{]}. Measurements remote from a lesion may therefore provide information about its location. However, this inference is complicated by the dependence of the hemodynamic response on anatomy, physiological driving conditions, and collateral pathways. For example, Yuhn et al. examined how
uncertainties in arterial diameters, stenosis parameters, and measured
flow rates affect predictions of collateral circulation and
postoperative flow redistribution {[}3{]}. Estimating lesion parameters from sparse waveforms therefore requires discrimination among candidate locations and severities, in addition to detecting abnormal velocities. Here, we investigate this discrimination under prescribed anatomical and physiological conditions.

Mechanistic blood-flow models relate vascular parameters to observable hemodynamic responses. One-dimensional models describe pulse-wave propagation along compliant vessels while coupling segments into an
arterial network. Reymond et al. evaluated systemic pressure and
flow predictions against human measurements {[}4{]}, while Boileau et al. compared numerical schemes using theoretical,
three-dimensional, and experimental benchmarks {[}5{]}. For the cerebral
circulation, Alastruey et al. studied how Circle of Willis
variations and occlusions affect flow {[}1{]}. Subsequent multiscale
approaches have broadened the anatomical and physiological description:
Hodneland et al. coupled large-vessel circulation with tissue
perfusion {[}6{]}, Ii et al. combined image-derived large
vessels with generated smaller vascular structures {[}7{]}, and Liu et al. incorporated cerebral autoregulation and pathway analysis to
investigate redistribution under anatomical variation and stenosis
{[}2{]}. These studies establish a modeling basis for relating a local
vascular perturbation to distributed responses. Their principal
objectives concern forward circulation, vascular structure, or
perfusion; recovering an unknown lesion from a restricted observation
set requires an additional inference procedure.

Model personalization has addressed part of this inverse problem by
estimating parameters that cannot be measured directly. Itu et al. developed a hierarchical procedure for tuning structured-tree
boundary conditions to patient-specific data {[}8{]}. Melis et al. used Gaussian-process emulators to facilitate sensitivity analysis of one-dimensional vascular models {[}9{]}. More recent
studies have examined which parameter combinations can be constrained:
Saxton et al. compared sensitivity-based parameter-selection
approaches in a lumped cardiovascular model {[}10{]}, and Mansilla
Alvarez et al. coupled global sensitivity analysis with
calibration of a one-dimensional circulation model from limited pressure
and flow measurements {[}11{]}. The distinction between sensitivity and identifiability is central to parameter estimation. A parameter may affect a measured
waveform without being uniquely recoverable when other parameters
produce similar changes. Accordingly, the present study formulates a conditional inverse problem: anatomy and physiological inputs are
treated as known, and the unknowns are the vessel containing a single
stenosis and its diameter reduction. This formulation isolates lesion discrimination from the broader problem of model personalization.

Physics-informed learning provides an approach to hemodynamic field reconstruction from incomplete observations. The PINN formulation of Raissi et al. incorporates differential-equation information into
neural-network training for forward and inverse problems {[}12{]}. In
cardiovascular applications, Kissas et al. combined non-invasive
flow measurements with one-dimensional equations to infer arterial
pressure {[}13{]}. Arzani et al. examined reconstruction of
near-wall flow from sparse data in idealized stenosis and aneurysm
geometries {[}14{]}. More directly related to cerebral networks,
Sarabian et al. combined transcranial Doppler velocities and
angiographic geometry with a physics-informed framework to reconstruct
pressure, area, and velocity, and compared predicted velocities with 4D
flow MRI {[}15{]}. They also demonstrated recovery of local caliber
changes in simulated cerebral vasospasm. Sen et al. subsequently
developed physics-informed graph neural networks for the one-dimensional
blood-flow equations {[}16{]}. Garay et al. investigated
simultaneous recovery of reduced-order model parameters and velocity
fields from sparse, noisy aortic measurements {[}17{]}. These studies demonstrate approaches to estimating unobserved quantities by combining
measurements and physical constraints. Their formulations address
different combinations of field reconstruction, boundary-parameter
estimation, and caliber changes. The present study poses an explicit
search over candidate lesion vessels and diameter-reduction levels under
known anatomical--physiological conditioning.

Waveform-based disease characterization provides a complementary approach to hemodynamic inference. Melis et al. studied cerebral vasospasm using a
one-dimensional circulation model and sensitivity analysis, identifying
waveform features associated with different narrowing conditions
{[}18{]}. Sen et al. developed a staged classification approach
to detect and localize thrombotic occlusions from simulated hemodynamic
waveforms {[}19{]}. Such studies connect disease-related changes to
observable signal structure, rather than treating reconstruction of an
entire field as the sole endpoint. Nevertheless, a classifier for an
occlusion category and a forward model conditioned on partial stenosis
address distinct inference tasks. Here, location is discrete, severity is
represented by diameter reduction, and both are evaluated through the
velocity response predicted for each candidate. The objective is to
retain a shared physical interpretation of localization, severity
estimation, and the associated reconstructed fields.

Surrogate models facilitate repeated evaluation of the forward problem. In cerebral circulation, Yuhn et al. trained
a neural surrogate of a 1D--0D model to support uncertainty
quantification of collateral flow {[}3{]}; this demonstrates the
relevance of learning a family of responses rather than solving only one
observed state. Operator learning extends this idea to mappings
involving input functions and solution fields {[}20{]}. DeepONet uses
branch and trunk networks to represent nonlinear operators {[}21{]}, the
Fourier neural operator parameterizes integral transformations in
Fourier space {[}22{]}, and the General Neural Operator Transformer
(GNOT) combines attention with geometric mixtures of experts to handle
heterogeneous inputs {[}23{]}. These constructions provide different
ways to share a learned forward map across conditions and query
locations. Their general capabilities do not, however, determine how a
localized lesion should be represented within a vascular network, or
whether its parameters are distinguishable from sparse observations.

A closely related operator-based cerebral reconstruction study is that of
Kaczmarek et al., who used Neural Operator Transformers with a
reduced-order model to reconstruct hemodynamics from sparse data and
infer boundary-related quantities {[}24{]}. Their inverse formulation
optimizes latent representations and global parameters using measurement
mismatch and physical constraints. This establishes an important
connection between operator surrogates and cerebral inverse problems.
Our emphasis is the explicit conditioning of a shared forward operator
on a localized stenosis, followed by comparison of candidate
vessel--severity pairs using observed velocity residuals. We distinguish
this task from both boundary-condition inference and direct waveform
classification. The comparison is methodological: the studies use
different data, unknown parameters, and evaluation protocols, so their
reported errors cannot establish a performance ranking for the present
problem.

This formulation imposes three modeling requirements. The representation must encode a localized lesion and its effects on other vessels, couple vessel-specific information to spatiotemporal field predictions through a shared model, and retain supervision of the healthy response when multiple lesion scenarios correspond to each baseline condition. We address these requirements with localized
lesion injection into vessel tokens, global self-attention across those
tokens, and severity-gated bilinear coupling to a shared
continuous-field decoder. A paired healthy-baseline objective directly
supervises the zero-severity path. The network is trained on reference
fields generated by a sequential 0D--1D simulator; the training objective does not include a differential-equation residual. During inverse estimation, the network weights remain fixed, and candidate residuals are evaluated using only the specified midpoint velocity observations.

The study contributes an explicit lesion-conditioned representation for
network-wide hemodynamic prediction (Section 4), a paired-baseline
training and candidate-residual inversion procedure (Section 5), and a
simulation-based assessment of how observation configuration and noise
affect location and severity recovery (Section 6). Forward field error,
lesion-localization accuracy, and severity error are evaluated
separately, because agreement in one endpoint does not establish
accuracy in the others. We also examine competing candidate residuals
and vessel-dependent failures to characterize ambiguity under the prescribed observation configuration. The scope is single-lesion recovery under known
anatomy and physiology within the stated scenario partition;
generalization to unseen anatomical conditions and clinical validation
remain separate questions.

\section{2. Mathematical model}\label{mathematical-model}

\subsection{2.1. Sequential 0D--1D hemodynamic
solver}\label{sequential-0d1d-hemodynamic-solver}

Hemodynamic fields computed using a sequentially coupled 0D--1D model served as the reference solutions for surrogate training and evaluation. For each anatomical--physiological condition, the 0D resistance
network determines the pressure and flow distribution used to initialize
the 1D compliant-vessel model and specify its boundary parameters. The
1D stage then resolves pulsatile propagation along the connected
vessels. The resulting fields of pressure \(P(x,t)\), cross-sectional area \(A(x,t)\), and cross-sectionally averaged velocity \(u(x,t)\) serve as supervised training targets for the neural operator.

Healthy and stenotic scenarios are solved within the same network and
boundary-condition framework. A lesion modifies the effective resistance
of its vessel and the reference lumen area within the stenotic segment, as
specified in Section 2.2. This formulation yields paired reference solutions with identical anatomical and physiological inputs and different lesion conditions.

\subsubsection{2.1.1. Zero-dimensional network
model}\label{zero-dimensional-network-model}

The 0D model represents vessel segments as hydraulic resistances connecting
internal, inlet, and terminal nodes. Periodic inlet pressure waveforms
are supplied by a precomputed cardiovascular circulation model, and
terminal nodes are assigned a reference pressure. For a vessel
connecting nodes \(i\) and \(j\), the resistance and nodal flow relations
use Poiseuille's law and the cerebral resistance-network representation
described by Kennedy McConnell and Payne {[}25{]}. The
Poiseuille resistance is

\[
R_{ij}=\frac{8\mu L_{ij}}{\pi r_{ij}^{4}},
\tag{1}
\]

where \(L_{ij}\) is centerline length, \(r_{ij}\) is mean radius, and
\(\mu\) is dynamic blood viscosity. The volumetric flow rate is related to the nodal pressure difference by

\[
Q_{ij}(t)=\frac{P_i(t)-P_j(t)}{R_{ij}}.
\tag{2}
\]

At each internal node, conservation of mass in the resistance network
{[}25{]} requires

\[
\sum_{j\in\mathcal N(i)}Q_{ij}(t)=0,
\tag{3}
\]

where \(\mathcal N(i)\) denotes the neighboring nodes. Equations
(1)--(3), together with the prescribed boundary pressures, define the nodal pressure system at each temporal sample. Segmental flow rates are then calculated from the nodal pressure differences, with their signs determining the flow direction in each vessel.

The 0D network is solved under a quasi-steady approximation at each temporal sample to determine nodal pressures and segmental flow directions. Spatially resolved pulsatile hemodynamics are subsequently computed using the 1D model.

\subsubsection{2.1.2. One-dimensional compliant-vessel
model}\label{one-dimensional-compliant-vessel-model}

Blood is treated as an incompressible fluid in an elastic tube. We use
the one-dimensional mass and momentum equations in area--velocity form
described by Sherwin et al. {[}26{]} and applied to the cerebral
circulation by Alastruey et al. {[}1{]}. The fields satisfy

\[
\frac{\partial A}{\partial t}+\frac{\partial Q}{\partial x}=0,
\qquad Q=uA,
\tag{4}
\]

and

\[
\frac{\partial u}{\partial t}
+\frac{\partial}{\partial x}\left(\frac{u^2}{2}+\frac{P}{\rho}\right)
=K_R\frac{u}{A},
\qquad K_R=-\frac{22\pi\mu}{\rho}.
\tag{5}
\]

Here, \(\rho\) denotes blood density and the right-hand side represents
viscous friction, with the \(22\pi\mu\) closure used in {[}1{]}.
We adopt the square-root functional form of the elastic pressure--area
law {[}26,1{]} and express it in terms of the reference wave speed as

\[
P-P_0=2\rho c_0^2\left(\sqrt{\frac{A}{A_0}}-1\right),
\tag{6}
\]

where \(A_0\) is reference area, \(P_0\) is reference pressure, and
\(c_0\) is the local reference wave speed. The Moens--Korteweg-type
expression, in the thin-wall convention stated by Filonova et al.
{[}27{]}, is

\[
c_0=\sqrt{\frac{Eh_0}{2\rho r_0}}.
\tag{7}
\]

Equation (7) specifies the wall-stiffness convention adopted here,
without the Poisson-ratio factor used in the stiffness parameterization
of Alastruey et al. {[}1{]}.

The constants are \(\rho=1060\ \mathrm{kg\,m^{-3}}\),
\(\mu=3.5\times10^{-3}\ \mathrm{Pa\,s}\), Young's modulus
\(E=4\times10^5\ \mathrm{Pa}\), and wall thickness
\(h_0=0.5\times10^{-3}\ \mathrm{m}\). The 0D nodal pressures and segmental flow rates initialize the 1D pressure and velocity fields and determine flow
directions and terminal resistance--compliance--resistance (RCR)
Windkessel parameters. The three-element terminal model follows the
RCR representation used in one-dimensional arterial modeling {[}4{]}.
The same numerical procedure is applied to the healthy and stenotic networks.

\subsection{2.2. Stenosis modeling}\label{stenosis-modeling}

We consider one lesion per diseased scenario. Following the
diameter-reduction convention used by Liang et al. {[}28{]}, severity is defined as

\[
s_l=1-\frac{D_s}{D_0},
\tag{8}
\]

where \(D_0\) and \(D_s\) are the normal and stenotic lumen diameters.
The stenosis is centered at the midpoint of the affected vessel. Within
its prescribed length, the reference area follows from Eq (8) and
the circular cross-section relation \(A=\pi D^2/4\):

\[
A_s=(1-s_l)^2A_0.
\tag{9}
\]

Both the prescribed lesion conditions and the estimated severities use
this diameter-reduction convention. An absolute error in severity is
therefore reported in percentage points of diameter reduction.

In the 0D resistance matrix, unaffected vessels retain their Poiseuille
resistances. The affected vessel is assigned a Young--Tsai-type effective resistance that accounts for additional viscous and inertial losses
{[}29,30{]}. Seeley and Young examined how stenosis geometry affects
pressure losses {[}31{]}. The present simulator uses the following additive resistance
convention, which differs from the original Young--Tsai pressure-drop law:

\[
R_{\mathrm{eff}}=R_{\mathrm{Pois}}+R_{\mathrm{visc}}+R_{\mathrm{inert}},
\tag{10}
\]

with

\[
\frac{R_{\mathrm{visc}}}{R_{\mathrm{Pois}}}
=\frac{0.83L_s+1.64D_s}{L}
\left(\frac{A_0}{A_s}\right)^2,
\tag{11}
\]

and

\[
R_{\mathrm{inert}}=
\frac{K_t\rho}{2A_0^2}
\left(\frac{A_0}{A_s}-1\right)^2|Q|,
\qquad K_t=1.52.
\tag{12}
\]

The effective-length coefficients \(0.83\) and \(1.64\), and the
contraction-loss coefficient \(K_t=1.52\), are given explicitly in the
stenosis pressure-drop parameterization of Liang et al. {[}28{]},
which uses the empirical model of Young and Tsai {[}30{]}.
Equation (11) normalizes its viscous resistance by Poiseuille resistance;
Eq (12) expresses its quadratic pressure loss as a flow-dependent
resistance. The acceleration-dependent term in that pressure-drop model
is omitted in the present quasi-steady update. Liu et al. {[}2{]}
provide a subsequent implementation of the resistance formulation in a
cerebral circulation model.

Here, \(L_s\) is stenosis length and \(L\) is the total length of the
affected vessel. The dependence of the inertial term on flow requires an iterative resistance update. The resistance is initialized from the lesion geometry and incorporated into the 0D network. The network is then solved, and the resulting flow rate is used to update the inertial contribution. These steps are repeated until the effective resistance converges.

After convergence of the resistance update, the nodal pressures and segmental flow rates are transferred to the 1D model. Stenosis is represented by a localized reduction in reference cross-sectional area over the prescribed interval. The compliant-vessel equations are then solved to compute the corresponding pulsatile hemodynamic fields. The reduced-order
lesion representation captures additional resistance and lumen
narrowing; three-dimensional jet structure and local wall stresses are not resolved.

\subsection{2.3. Dataset construction}\label{dataset-construction}

Reference simulations were performed for lesion scenarios nested within sampled anatomical and physiological conditions. Latin
hypercube sampling {[}32{]} varied heart rate, diastolic and systolic
inlet pressures, and radius scales for the 20 target arteries. Within
each sampled condition, scenarios differed in the affected vessel and
diameter reduction, with an additional healthy solution providing the
matched baseline. Table 1 summarizes the sampling ranges and lesion
settings.

\textbf{Table 1.} Parameters used to construct the simulated stenosis
dataset. Anatomy--physiology parameters vary between virtual
conditions; lesion parameters vary between scenarios within a condition.
Severity is diameter reduction.
{\def\LTcaptype{none} 
\begin{longtable}[]{@{}
  >{\raggedright\arraybackslash}p{(\linewidth - 6\tabcolsep) * \real{0.2500}}
  >{\raggedright\arraybackslash}p{(\linewidth - 6\tabcolsep) * \real{0.2500}}
  >{\raggedright\arraybackslash}p{(\linewidth - 6\tabcolsep) * \real{0.2500}}
  >{\raggedright\arraybackslash}p{(\linewidth - 6\tabcolsep) * \real{0.2500}}@{}}
\toprule\noalign{}
\begin{minipage}[b]{\linewidth}\raggedright
Parameter
\end{minipage} & \begin{minipage}[b]{\linewidth}\raggedright
Variation level
\end{minipage} & \begin{minipage}[b]{\linewidth}\raggedright
Range or values
\end{minipage} & \begin{minipage}[b]{\linewidth}\raggedright
Setting
\end{minipage} \\
\midrule\noalign{}
\endhead
\bottomrule\noalign{}
\endlastfoot
Heart rate & Anatomy--physiology & 60--80 beats/min & Latin hypercube
sampling \\
Diastolic pressure & Anatomy--physiology & 9.87--11.47 kPa & Latin
hypercube sampling \\
Systolic pressure & Anatomy--physiology & 15.3--18.0 kPa & Latin
hypercube sampling \\
Radius scale & Anatomy--physiology & 0.7--1.3 & 20 target arteries \\
Stenotic vessel & Scenario & 20 candidate arteries & Single lesion \\
Diameter reduction & Scenario & 0.10, 0.20, 0.35, 0.40, 0.45, 0.60, 0.70
& Prescribed severity values \\
Healthy baseline & Scenario & \(s_l=0\) & One per condition \\
Lesion center & Fixed & Vessel midpoint & All diseased scenarios \\
Lesion length & Fixed & 1 cm & All diseased scenarios \\
Spatial sampling & Fixed & 50 points per vessel & Resampled fields \\
Temporal sampling & Fixed & 100 points per cardiac cycle & Resampled
fields \\
\end{longtable}
}

The dataset comprises 31 anatomical--physiological conditions. For each
scenario, the simulated fields were resampled at 50 axial locations per
vessel and 100 time points over one cardiac cycle. This sampling yields 5,000 query locations per vessel, with pressure, area, and velocity defined at
each location. The same sampling convention was used when comparing
predicted and reference fields.

Within each anatomical--physiological condition, scenarios were assigned
to training, validation, and test subsets in approximate proportions of
70\%, 15\%, and 15\%. Each scenario belonged to one subset, whereas the
underlying anatomical--physiological conditions were shared across
subsets. Evaluation therefore concerns held-out lesion scenarios within
the sampled conditions, rather than prediction for previously unseen
anatomy.

Healthy fields matched to training scenarios were also used for the
auxiliary baseline objective in Section 5.2. These fields are part of
the training supervision. The distinction between conditions and
scenarios is retained throughout the evaluation: the 31 sampled
conditions define anatomical diversity, whereas additional vessels, time
points, and lesion states provide observations within those conditions.

\section{4. Neural network framework}\label{neural-network-framework}

\subsection{4.1. Neural forward
surrogate}\label{neural-forward-surrogate}

Using the reference dataset, we trained a shared neural operator to map
anatomical--physiological inputs and lesion parameters to pulsatile
hemodynamic fields. The operator is evaluated separately for each target vessel, with model parameters shared across vessels. For vessel \(i\),
let \(\Omega_i\) denote the axial domain and \(\mathcal T\) the
cardiac-cycle interval. At a query
\(\xi=(x,t)\in\Omega_i\times\mathcal T\), the output field is

\[
y_i(\xi)=[P_i(\xi),A_i(\xi),u_i(\xi)].
\tag{13}
\]

We define the lesion-conditioned forward map used in this study as

\[
\widehat y_i(\xi)=\mathcal F_\theta(\xi,A_{0,i},p_i,v_l,s_l),
\qquad i=1,\ldots,N_a,
\tag{14}
\]

where \(N_a=20\), \(A_{0,i}\) is the reference area function, and
\(p_i=[e_i,r,g]\) contains the target-vessel identity \(e_i\), the
vector of all target-vessel radii \(r\), and global physiological inputs
\(g\). The lesion is specified by vessel index \(v_l\) and severity
\(s_l\).

The architecture separates network-level lesion encoding from
within-vessel field decoding. The vessel-token branch produces a
lesion-conditioned feature for every artery. This feature modulates a
coordinate-based decoder that also receives the target vessel identity,
anatomical--physiological inputs, and reference area. The following
sections describe the token interaction and field decoder in the order
shown in Fig 1.

\subsection{4.2. Vessel-token
propagation}\label{vessel-token-propagation}

\subsubsection{4.2.1. Context encoding and localized lesion
injection}\label{context-encoding-and-localized-lesion-injection}

We define the following vessel-token representation for this study.
The propagation module produces a matrix of vessel-specific features,

\[
H^{\mathrm{prop}}=\mathcal P_\phi(v_l,s_l,r,g)
\in\mathbb R^{N_a\times d_p}.
\tag{15}
\]

Its \(i\)th row, \(h_i^{\mathrm{prop}}\), represents the
lesion-conditioned context for target vessel \(i\). Before introducing a
lesion, each vessel is assigned a token

\[
c_i=e_i^{\mathrm{art}}+E_{\mathrm{ctx}}(r_i,g),
\tag{16}
\]

where \(e_i^{\mathrm{art}}\) is a learned identity embedding and
\(E_{\mathrm{ctx}}\) is shared across vessels. The local radius and
global physiological inputs therefore affect the token alongside vessel
identity.

A shared encoder maps lesion severity to a source feature, which is
inserted only into the affected vessel:

\[
q_l=s_lE_{\mathrm{sten}}(s_l),
\qquad h_i^{(0)}=c_i+\mathbb I(i=v_l)q_l.
\tag{17}
\]

Stacking the vessel tokens forms the initial sequence \(H^{(0)}\). The explicit lesion feature is initially restricted to the affected vessel and is zero when \(s_l=0\). Subsequent token interactions incorporate the lesion condition into the network representation.

\subsubsection{4.2.2. Cross-vessel
interaction}\label{cross-vessel-interaction}

The token sequence passes through \(L_p\) Transformer encoder layers
{[}33{]}:

\[
H^{(m+1)}=\mathcal T_{\phi_m}(H^{(m)}),
\qquad m=0,\ldots,L_p-1,
\tag{18}
\]

and \(H^{\mathrm{prop}}=H^{(L_p)}\). Each layer contains multi-head
self-attention and a token-wise feed-forward transformation, with
residual connections and layer normalization. Repeated interaction
allows the feature associated with a local lesion to modify
representations of other vessels in a manner conditioned on the input
anatomy and physiology.

The propagation branch permits interactions among all vessel tokens without an anatomical adjacency mask. Here, propagation refers to learned interactions between vessel representations. The physical network remains part of the
reference solver; attention weights in the surrogate are not prescribed
flow fractions.

\subsection{4.3. Stenosis-conditioned continuous-field
operator}\label{stenosis-conditioned-continuous-field-operator}

\subsubsection{4.3.1. Query representation and bilinear
coupling}\label{query-representation-and-bilinear-coupling}

For a target vessel, the query coordinates and finite-dimensional
anatomical--physiological condition are encoded by a shared trunk:

\[
z_i^{\mathrm{base}}(\xi)=E_q(\xi,p_i).
\tag{19}
\]

The lesion index and severity do not directly enter this trunk
representation. They are encoded by the propagation branch and
introduced through the severity-gated bilinear interaction defined here,

\[
z_i^{(0)}(\xi)=z_i^{\mathrm{base}}(\xi)
+s_l\mathcal B\big(z_i^{\mathrm{base}}(\xi),h_i^{\mathrm{prop}}\big),
\tag{20}
\]

where

\[
\mathcal B(z,h)=W_c[(U_cz)\odot(V_ch)].
\tag{21}
\]

The learned maps \(U_c\), \(V_c\), and \(W_c\) define the bilinear
transformation, and \(\odot\) denotes element-wise multiplication. The vessel-level feature is shared across all query locations within that vessel,
but its interaction with \(z_i^{\mathrm{base}}(\xi)\) allows spatial and temporal variation in the modulation.

At \(s_l=0\), the explicit modulation in Eq (20) vanishes, leaving the
baseline query representation. The healthy-baseline objective in Section
5.2 directly supervises this path. For nonzero severity, the encoder,
token interaction, and decoder retain a nonlinear dependence on the
lesion condition.

\subsubsection{4.3.2. Geometry encoding and GNOT
blocks}\label{geometry-encoding-and-gnot-blocks}

The reference area function is encoded independently as

\[
a_i=E_A(A_{0,i}).
\tag{22}
\]

The conditioned query sequence is processed by \(L\) shared GNOT blocks,
using normalized linear attention and geometric mixtures of experts
{[}23{]}. Cross-attention first incorporates the geometry sequence:
queries originate from the current field representation, and keys and
values originate from \(a_i\). The normalized linear formulation
rearranges the attention calculation to avoid explicitly constructing
the full query--key interaction matrix. For fixed feature width, this
reduces the growth of attention cost with the number of queries.

After cross-attention, a coordinate-dependent gate combines the expert
transformations. Following the geometric gating construction of GNOT
{[}23{]}, we write the expert weights in block \(l\) as

\[
\pi_k^{(l)}(\xi)=
\frac{\exp G_k^{(l)}(\xi)}{\sum_{q=1}^{K}\exp G_q^{(l)}(\xi)},
\qquad k=1,\ldots,K,
\tag{23}
\]

and adapt its weighted expert aggregation {[}23{]} to two expert groups:

\[
\mathcal M_j^{(l)}(z,\xi)=
\sum_{k=1}^{K}\pi_k^{(l)}(\xi)E_{k,j}^{(l)}(z),
\qquad j\in\{1,2\}.
\tag{24}
\]

We use \(K=4\) experts. The first expert group follows cross-attention.
Normalized linear self-attention then models dependencies among query
locations within the target vessel, followed by the second expert group.
The two groups have independent expert parameters but share the
coordinate-derived gating weights within a block. Residual connections
and layer normalization are used throughout.

After the final block, a shared output multilayer perceptron maps the
hidden representation to the three physical channels,

\[
\widehat y_i(\xi)=E_{\mathrm{out}}(z_i^{(L)}(\xi))
=[\widehat P_i(\xi),\widehat A_i(\xi),\widehat u_i(\xi)].
\tag{25}
\]

The output map is shared across arteries, with vessel identity,
geometry, physiology, and the propagated lesion feature specifying each
response. Self-attention couples locations within the supplied query
sequence. We therefore retain the query-grid convention during
evaluation; coordinate-based outputs alone do not establish invariance
to changes in that grid.

\section{5. Training strategy and inverse
estimation}\label{training-strategy-and-inverse-estimation}

\subsection{5.1. Normalization}\label{normalization}

Continuous input variables and output channels are standardized using
training-set statistics,

\[
\widetilde z=\frac{z-\mu_z}{\sigma_z}.
\tag{26}
\]

The mean \(\mu_z\) and standard deviation \(\sigma_z\) are calculated
from the training data and then fixed. Validation and test data use the
same statistics. The discrete lesion-vessel index, vessel one-hot
identity, and continuous stenosis severity retain their original
representations. Thus, a severity of zero continues to identify the
healthy state directly.

Training losses are evaluated in standardized output space so that
pressure, area, and velocity, which differ in physical units and magnitudes, are represented on a common numerical scale. Predicted fields are
transformed back to physical units for waveform comparison and
channel-specific error reporting.

\subsection{5.2. Field reconstruction and paired baseline
anchoring}\label{field-reconstruction-and-paired-baseline-anchoring}

Let \(Y_{b,i}\) and \(\widehat Y_{b,i}\) denote the reference and
predicted field matrices for scenario \(b\) and artery \(i\) in
standardized space. The reconstruction objective is the mean relative
Frobenius error,

\[
\mathcal L_{\mathrm{field}}=
\frac{1}{BN_a}\sum_{b=1}^{B}\sum_{i=1}^{N_a}
\frac{\|\widehat Y_{b,i}-Y_{b,i}\|_F}
{\max(\|Y_{b,i}\|_F,\varepsilon)},
\tag{27}
\]

where \(B\) is minibatch size and \(\varepsilon\) prevents division by a
vanishing norm. Both healthy and diseased scenarios contribute to the
field objective.

The dataset contains multiple stenotic scenarios for each healthy condition. Uniform weighting of individual scenarios therefore assigns a smaller proportion of direct supervision to the healthy state. To increase supervision of the zero-severity response, each training scenario is paired with the healthy
field from the same anatomical--physiological condition. With the lesion
condition set to \(s_l=0\), the same network produces
\(\widehat Y_{b,i}^{(0)}\). The baseline objective is

\[
\mathcal L_{\mathrm{base}}=
\frac{1}{BN_a}\sum_{b=1}^{B}\sum_{i=1}^{N_a}
\frac{\|\widehat Y_{b,i}^{(0)}-Y_{b,i}^{(0)}\|_F}
{\max(\|Y_{b,i}^{(0)}\|_F,\varepsilon)}.
\tag{28}
\]

The total loss is

\[
\mathcal L_{\mathrm{total}}=
\mathcal L_{\mathrm{field}}+\lambda_{\mathrm{base}}\mathcal L_{\mathrm{base}},
\qquad \lambda_{\mathrm{base}}=0.5.
\tag{29}
\]

This auxiliary term supervises the paired healthy field, rather than the
difference between healthy and stenotic fields. It complements the
zero-severity path in Eq (20) by providing direct baseline supervision
for each training condition. Algorithm 1 summarizes the two forward
passes and the shared parameter update. In the algorithm, \(b\) indexes
scenarios, \(i\) indexes target vessels, and \(\Xi_i\) is the supplied
query grid; the notation \(\mathcal F_\theta(\Xi_i,\ldots)\) denotes
evaluation at all queries in that vessel.

\begin{nolinenumbers}
\begin{algorithm}[H]
\caption{Training with paired healthy-baseline anchoring}
\label{alg:algorithm1}
\small\setstretch{1.08}
\begin{algorithmic}[1]
\Require Training and validation sets \(\mathcal D_{\mathrm{tr}},\mathcal D_{\mathrm{val}}\); paired healthy fields; query grid \(\Xi\); number of epochs \(E\).
\Ensure Trained parameters \(\theta\) and fixed normalization statistics \(\mathcal S\).
\State Fit normalization statistics \(\mathcal S\) using \(\mathcal D_{\mathrm{tr}}\) only.
\State Initialize the shared surrogate parameters \(\theta\).
\For{\(e=1,\ldots,E\)}
\For{each minibatch \(\mathcal B\subset\mathcal D_{\mathrm{tr}}\)}
\State Standardize continuous inputs and fields using \(\mathcal S\).
\State \(\widehat Y_{b,i}\leftarrow \mathcal F_\theta(\Xi_i,A_{0,b,i},p_{b,i},v_b,s_b)\) for all \((b,i)\).
\State \(\widehat Y_{b,i}^{(0)}\leftarrow \mathcal F_\theta(\Xi_i,A_{0,b,i},p_{b,i},v_b,0)\) for all \((b,i)\).
\State Evaluate relative Frobenius loss \(\mathcal L_{\mathrm{field}}\) using Eq (27).
\State Evaluate paired healthy-field loss \(\mathcal L_{\mathrm{base}}\) using Eq (28).
\State \(\mathcal L\leftarrow\mathcal L_{\mathrm{field}}+0.5\mathcal L_{\mathrm{base}}\).
\State Update \(\theta\) with AdamW using \(\nabla_\theta\mathcal L\) and the cosine schedule.
\EndFor
\State Evaluate the surrogate on \(\mathcal D_{\mathrm{val}}\) using the fixed statistics \(\mathcal S\).
\EndFor
\State \Return \(\theta,\mathcal S\).
\end{algorithmic}
\end{algorithm}
\end{nolinenumbers}

\subsection{5.3. Inversion from sparse velocity
waveforms}\label{inversion-from-sparse-velocity-waveforms}

After training, lesion parameters are estimated by minimizing the mismatch between candidate forward predictions and sparse velocity observations. The
anatomical--physiological inputs are fixed, and the unknowns are lesion
vessel \(v_l\) and severity \(s_l\). Measurements are velocity waveforms
at the midpoint of vessels in the observation set \(\mathcal O\). For
each candidate \((v,s)\in\mathcal C\), the surrogate predicts the
hemodynamic fields and extracts the velocities at those observation
sites.

The selected lesion parameters minimize the mean squared waveform
discrepancy,

\[
(\widehat v_l,\widehat s_l)=
\operatorname*{arg\,min}_{(v,s)\in\mathcal C}
\frac{1}{N_{\mathrm{obs}}N_t}
\sum_{i\in\mathcal O}\sum_{j=1}^{N_t}
\left[\widehat u_i(x_{\mathrm{mid}},t_j;v,s)
-u_i^{\mathrm{obs}}(x_{\mathrm{mid}},t_j)\right]^2.
\tag{30}
\]

Here, \(N_{\mathrm{obs}}\) is the number of observed vessels and \(N_t\)
is the number of temporal samples. The surrogate weights remain fixed
throughout candidate search. The residual is computed using only the observed and predicted midpoint velocities; pressure, area, and full-field reference solutions are
reserved for forward evaluation. Algorithm 2 summarizes the candidate evaluation procedure.

\begin{nolinenumbers}
\begin{algorithm}[H]
\caption{Stenosis inversion by candidate waveform residuals}
\label{alg:algorithm2}
\small\setstretch{1.08}
\begin{algorithmic}[1]
\Require Fixed surrogate \(\mathcal F_\theta\) and statistics \(\mathcal S\); known anatomy, physiology and reference areas; query grid \(\Xi\); candidates \(\mathcal C\); observed velocities \(u^{\mathrm{obs}}\) at vessels \(\mathcal O\).
\Ensure Estimated lesion \((\widehat v_l,\widehat s_l)\) and reconstructed fields \(\widehat Y\).
\State Standardize continuous model inputs using \(\mathcal S\); keep \(\theta\) fixed.
\For{each candidate \((v,s)\in\mathcal C\)}
\State \(\widehat Y_i^{(v,s)}\leftarrow \mathcal F_\theta(\Xi_i,A_{0,i},p_i,v,s)\) for all target vessels \(i\).
\State Transform predicted fields to physical units using \(\mathcal S\).
\State Extract \(\widehat u_i^{(v,s)}(x_{\mathrm{mid}},t_j)\) for \(i\in\mathcal O\) and \(j=1,\ldots,N_t\).
\State Compute \(J(v,s)\) as the observed-velocity MSE in Eq (30).
\State Store \(J(v,s)\) and the corresponding predicted fields \(\widehat Y^{(v,s)}\).
\EndFor
\State \((\widehat v_l,\widehat s_l)\in\operatorname*{arg\,min}_{(v,s)\in\mathcal C}J(v,s)\).
\State \Return \((\widehat v_l,\widehat s_l)\) and \(\widehat Y^{(\widehat v_l,\widehat s_l)}\).
\end{algorithmic}
\end{algorithm}
\end{nolinenumbers}

\subsection{5.4. Evaluation measures}\label{evaluation-measures}

Forward reconstruction is evaluated with the true lesion condition
supplied to the model. For a physical channel \(q\in\{P,A,u\}\) and
target vessel \(i\), the relative error is

\[
E_{i,q}=100\,
\frac{\|\widehat q_i-q_i\|_2}{\|q_i\|_2}.
\tag{31}
\]

Equation (31) quantifies the relative reconstruction error for each physical variable over the evaluation spatiotemporal grid. A small reference norm can increase the relative
error even when the absolute discrepancy is modest, so vessel-level
errors are interpreted together with representative waveforms.

Localization is measured by Top-1 accuracy, the proportion of evaluated
lesion scenarios for which the selected vessel equals the true vessel.
Severity estimation is assessed using mean absolute error (MAE) in diameter
reduction. Under the stated analysis protocol, the principal
severity error is evaluated conditional on correct localization.
Percentages used for diameter reduction correspond to percentage points
when an absolute severity difference is reported. Localization accuracy and severity error are evaluated as distinct performance measures.

\section{6. Results}\label{results}

\subsection{6.1. Experimental setup}\label{experimental-setup}

The model was trained for 200 epochs on one NVIDIA H100 GPU using the
AdamW optimizer {[}34{]}, a cosine learning-rate schedule, and a batch
size of one. The evaluation comprised forward prediction with prescribed lesion parameters and inverse estimation from sparse midpoint velocity observations. The
reference solutions were generated by the 0D--1D solver. Four
observation configurations, denoted 4v, 6v2, 8v, and 8v2, were compared
using the configuration labels defined for the study. Localization was evaluated
over all 20 candidate vessels and over the designated KEY-11 subset.

Forward and inverse evaluations used the scenario-based partition
described in Section 3.2. Table 2 reports the forward-validation
summary. The inverse evaluation then assessed lesion localization and
severity under the specified observation configurations and noise
settings. This evaluation distinguishes forward approximation error from errors in lesion parameter estimation.

\subsection{6.2. Full-field reconstruction
performance}\label{full-field-reconstruction-performance}

With the lesion parameters supplied, the surrogate reproduced the
reference pressure, area, and velocity fields with reported
network-average relative errors of 0.14\%, 0.59\%, and 2.79\%,
respectively (Table 2). Representative temporal profiles are shown in
Fig 2. The mean relative error was higher for velocity than for pressure or area. Vessel-specific velocity errors were therefore examined before inverse localization.

\textbf{Table 2.} Forward relative \(L_2\) errors in the
validation summary. Vessel pairs are reported in the order indicated.
The vessel rows show the available artery-level breakdown; the overall
row gives the reported network summary and is not recalculated from the
displayed subset.
{\def\LTcaptype{none} 
\begin{longtable}[]{@{}
>{\raggedright\arraybackslash}p{(\linewidth - 6\tabcolsep) * \real{0.34}}
>{\raggedleft\arraybackslash}p{(\linewidth - 6\tabcolsep) * \real{0.24}}
>{\raggedleft\arraybackslash}p{(\linewidth - 6\tabcolsep) * \real{0.195}}
>{\raggedleft\arraybackslash}p{(\linewidth - 6\tabcolsep) * \real{0.225}}@{}}
\toprule\noalign{}
\begin{minipage}[b]{\linewidth}\raggedright
Vessel or vessel pair
\end{minipage} & \begin{minipage}[b]{\linewidth}\raggedleft
Pressure error (\%)
\end{minipage} & \begin{minipage}[b]{\linewidth}\raggedleft
Area error (\%)
\end{minipage} & \begin{minipage}[b]{\linewidth}\raggedleft
Velocity error (\%)
\end{minipage} \\
\midrule\noalign{}
\endhead
\bottomrule\noalign{}
\endlastfoot
RICA / LICA & 0.14 / 0.14 & 0.29 / 0.25 & 1.14 / 0.91 \\
RVA / LVA & 0.18 / 0.18 & 0.63 / 0.63 & 1.00 / 1.54 \\
BA & 0.15 & 0.47 & 0.58 \\
RMCA1 / LMCA1 & 0.13 / 0.13 & 0.64 / 0.63 & 0.54 / 0.53 \\
RACA1 / LACA1 & 0.13 / 0.12 & 0.59 / 0.57 & 4.85 / 4.03 \\
RPCA1 / LPCA1 & 0.14 / 0.13 & 0.84 / 0.83 & 2.29 / 2.06 \\
ACoA & 0.13 & 0.75 & 12.14 \\
RPCoA / LPCoA & 0.13 / 0.14 & 0.71 / 0.95 & 7.00 / 13.65 \\
Reported network average & 0.14 & 0.59 & 2.79 \\
\end{longtable}
}

The vessel-level distribution was markedly heterogeneous (Fig 3). Across
the 14 arteries listed in Table 2, pressure errors ranged from 0.12\% to
0.18\%, and area errors ranged from 0.25\% to 0.95\%. Velocity errors
spanned a wider range, from 0.53\% in LMCA1 to 13.65\% in LPCoA. The
communicating arteries had the three largest listed velocity errors:
LPCoA, 13.65\%; ACoA, 12.14\%; and RPCoA, 7.00\%. In contrast, both MCA1
segments and BA had errors below 0.6\%. These comparisons use the
displayed vessel entries; the reported network averages were retained
without recalculation from that incomplete subset.

Bilateral comparisons further showed that the error pattern depended on
both vessel group and side. Velocity errors were similar in RMCA1 and
LMCA1 (0.54\% and 0.53\%), whereas RPCoA and LPCoA differed by 6.65
percentage points (7.00\% and 13.65\%). The corresponding pressure
errors remained close, at 0.13\% and 0.14\%. Thus, comparable pressure errors across vessels did not imply comparable velocity reconstruction accuracy. In the
representative waveform example, predicted profiles followed the
reference cardiac-cycle variation under the true lesion condition. This
comparison measures forward approximation and excludes errors caused by
selecting an incorrect lesion in the inverse search.

\subsection{6.3. Inverse localization and severity
estimation}\label{inverse-localization-and-severity-estimation}

We next estimated the lesion parameters using only the specified
midpoint velocity waveforms. Without added noise, All-20 localization
accuracy was 96.6\% for 4v, 99.2\% for 6v2, 94.8\% for 8v, and 99.4\%
for 8v2 (Table 3). The 8v2 configuration achieved the highest localization accuracy among the evaluated configurations. Localization was therefore not determined by
observation count alone: the six-vessel 6v2 configuration also achieved higher accuracy than the eight-vessel 8v configuration.

\textbf{Table 3.} Top-1 stenosis-localization accuracy (\%) under the
study's TCD velocity-noise settings. Observation configurations are
denoted 4v, 6v2, 8v, and 8v2. KEY-11 is a subset of candidate lesion
vessels, not the number of observed vessels. The percentages are
empirical accuracies for the evaluated scenarios.
{\def\LTcaptype{none} 
\begin{longtable}[]{@{}llrrrr@{}}
\toprule\noalign{}
Target subset & Obs. & 0\% noise & 10\% noise & 20\% noise & 30\%
noise \\
\midrule\noalign{}
\endhead
\bottomrule\noalign{}
\endlastfoot
All 20 vessels & 4v & 96.6 & 95.7 & 93.5 & 90.3 \\
All 20 vessels & 6v2 & 99.2 & 97.7 & 96.5 & 95.8 \\
All 20 vessels & 8v & 94.8 & 94.0 & 94.8 & 91.7 \\
All 20 vessels & 8v2 & 99.4 & 98.8 & 97.5 & 96.0 \\
KEY-11 vessels & 4v & 99.4 & 98.9 & 99.4 & 98.1 \\
KEY-11 vessels & 8v2 & 100.0 & 100.0 & 100.0 & 100.0 \\
\end{longtable}
}

For 8v2 without added noise, most severity estimates lay near the
identity line (Fig 4), with a reported MAE of 1.8 percentage points
conditional on correct vessel localization. The scatter also
showed substantial severity errors among points marked as
correctly localized. For example, at a true diameter reduction of 60\%,
correctly localized points extended both below and above the identity
line. These results distinguish vessel localization from estimation of the degree of narrowing within the identified artery.

The severity scatter showed deviations in both directions rather than a
uniform overestimation or underestimation across the displayed range.
Substantial errors in individual cases persisted despite the low aggregate MAE. Because the scatter includes both correct- and
wrong-vessel markers, the scatter distribution must be interpreted alongside the severity statistic conditional on correct localization and the independently reported localization accuracy.

\subsection{6.4. Effect of sparse observation
configurations}\label{effect-of-sparse-observation-configurations}

Observation placement affected inversion at a fixed measurement count.
Both 8v and 8v2 used eight observed vessels, yet 8v2 achieved higher
All-20 localization accuracy at every reported noise level. The
difference was 4.6 percentage points without added noise and 4.3
percentage points at the 30\% setting. Thus, measurement count alone did
not explain the ranking of the observation configurations.

The benefit of observation placement persisted under noise. Relative to
8v, the 6v2 configuration improved All-20 localization by 4.4, 3.7, 1.7,
and 4.1 percentage points at the 0\%, 10\%, 20\%, and 30\% settings,
respectively. The corresponding 8v2 improvements were 4.6, 4.8, 2.7, and
4.3 percentage points. These differences are calculated directly from
Table 3. Across all four settings, an observation configuration with
fewer vessels could outperform one with more vessels, while changing the
placement of eight observations consistently changed the attainable
accuracy.

Localization was also dependent on the candidate-vessel subset. For 4v,
KEY-11 accuracy ranged from 98.1\% to 99.4\%, compared with 90.3\% to
96.6\% for All-20. The difference between KEY-11 and All-20 accuracy widened
from 2.8 percentage points without noise to 7.8 percentage points at the
30\% setting. Under 8v2, all evaluated KEY-11 cases were correctly
localized at every reported setting, while All-20 accuracy fell from
99.4\% to 96.0\%. These subset comparisons locate the remaining failures
outside KEY-11 and explain why its perfect empirical accuracy does not
summarize the full candidate space. Fig 5 presents the
observation-configuration and noise trends.

\subsection{6.5. Robustness to observation
noise}\label{robustness-to-observation-noise}

Increasing observation noise generally reduced localization accuracy,
with different sensitivities across configurations. The All-20 accuracy
for 4v decreased by 6.3 percentage points between the zero-noise and
30\% settings, compared with decreases of 3.4, 3.1, and 3.4 percentage
points for 6v2, 8v, and 8v2. Small nonmonotonic changes, such as the 8v
result at 20\% noise, were present in the finite evaluation and should
not be interpreted as a beneficial effect of noise.

Severity MAE and localization accuracy exhibited different trends
(Fig 5). The 8v2 configuration had the lowest displayed severity MAE at
each noise setting, whereas 4v had the highest. The 6v2 curve rose at
20\% noise and then decreased at 30\%, despite a continued decrease in
its localization accuracy. These differing trends indicate that the two metrics characterize distinct aspects of parameter recovery and do not establish improved performance with additional noise.

Stratification by lesion severity revealed additional variation in localization accuracy (Fig 6). Mild stenoses had the lowest localization accuracy in
each noise panel, and the reduction at 30\% noise was most apparent for
4v. Moderate and severe groups generally retained higher localization
accuracy, although their ordering was not uniform across configurations.
The aggregate noise curves therefore reflect observation-configuration effects
with differences in lesion severity. Noise percentages here denote the
prescribed simulation settings, and the plotted summaries are not
population-level confidence intervals.

\subsection{6.6. Representative errors and vessel-dependent
distinguishability}\label{representative-errors-and-vessel-dependent-distinguishability}

Vessel-resolved analysis showed that errors were concentrated in
selected communicating and distal arteries rather than distributed
evenly across the network (Fig 7). The vessel-resolved heatmap pools six
observation configurations and extends the perturbations to the 50\% and
80\% stress-test settings; its percentages are therefore distinct from
the four-configuration comparisons in Table 3. Within this pooled
visualization, mislocalization rates increased markedly for several distal vessels as noise increased, whereas rates for other vessels remained close to zero.
The heatmap therefore identified vessel-specific responses to noise that were not apparent from aggregate localization accuracy.

The residual-margin plots compared the separation of the noiseless
candidate minimum with mislocalization at the 30\% and 80\% settings. A
shallow residual minimum was associated with elevated error in several
vessels, but the relation was not uniform: some vessels exhibited appreciable mislocalization despite larger margins. At the higher perturbation level, changes in mislocalization rates varied across vessels. These plots
therefore support a descriptive association between candidate separation
and localization difficulty, without reducing all failures to a single
margin threshold.

The margin analysis concerns the finite candidate search used here. Its
interpretation depends on the surrogate, candidate grid, known
conditioning inputs, and observation configuration. It characterizes
practical discrimination within that setup rather than structural
identifiability of the continuous hemodynamic model.

\subsection{6.7. Learned inter-vessel
interactions}\label{learned-inter-vessel-interactions}

The aggregated attention summary showed nonuniform interactions among
vessel tokens (Fig 8). Several bilateral pairs, including the vertebral
arteries and paired ACA segments, formed visually prominent off-diagonal
entries. Other strong entries occurred across vessel groups, while the
displayed anatomical-edge and bilateral-pair overlays did not account
for every high-weight location. The learned representation therefore
combined local and nonlocal token associations.

These patterns were obtained without an explicit anatomical attention
mask. The heatmap is a symmetrized, aggregated representation of
attention, so it does not preserve directional flow information or
quantify blood-flow redistribution. This analysis characterizes the interactions learned by the propagation branch and complements the evaluation of forward reconstruction and inverse residuals.

\section{7. Discussion}\label{discussion}

The central finding is that a shared, lesion-conditioned forward
operator can support conditional localization and severity estimation
from sparse cerebral velocity waveforms. The distinction between forward
and inverse endpoints is essential to interpreting this result. Small
pressure and area errors showed that the operator reproduced important
features of the reference hemodynamics, but velocity reconstruction accuracy varied substantially across arteries. Inverse estimation accuracy also depended on the measurement locations: 6v2 exceeded 8v at every reported noise
setting, and 8v2 consistently exceeded 8v despite having the same
measurement count. Together, these findings indicate that mean field accuracy and the number of observed vessels are insufficient to characterize lesion discrimination.

These findings are consistent with hemodynamic coupling within the cerebral arterial network. Circle of Willis models show that local obstruction can
alter flow through communicating pathways {[}1,2{]}, while uncertainty
analysis has demonstrated the influence of vessel diameters and stenosis
conditions on collateral-flow predictions {[}3{]}. Measurements can therefore contain information about remote lesions, although distinct lesion hypotheses may produce similar responses at a restricted set of measurement sites. The present comparison of observation configurations provides computational
evidence for this dependence under known anatomy and physiology. It
motivates selecting measurements according to their ability to separate
clinically relevant lesion hypotheses, rather than assuming that every
additional vessel contributes equivalent information. Such selection
remains to be tested prospectively on independent conditions.

The largest listed forward velocity errors occurred in the communicating
arteries, whereas pressure errors remained relatively uniform. This disparity is relevant to the interpretation of both surrogate accuracy and inverse estimation errors. Relative velocity error can be amplified when the
reference flow is small, and the waveform mismatch relevant to
localization is evaluated only at the observed sites. Consequently, an
artery with a large normalized forward error need not be the artery most
difficult to localize, and a small network-average error cannot exclude
a poorly separated candidate pair. The residual-margin analysis
addresses the latter question more directly, although its exceptions
show that no single margin statistic fully accounts for the observed
vessel-level failures. Absolute errors, waveform shape, observation
placement, and candidate separation should be examined together in
subsequent validation.

The architecture represents localized lesions and their distributed hemodynamic effects. A lesion feature is inserted only
into the affected vessel token, then shared across the latent network
representation through global self-attention. Severity-gated bilinear
coupling allows that context to modify the spatial and temporal response
of each target vessel, while the paired healthy objective supervises the
background state under the same anatomical--physiological condition. The
attention map demonstrates that the trained branch used nonuniform
cross-vessel associations. It does not show recovery of a physical flow
graph: the summary is symmetrized, attention is not a flow fraction, and
no perturbation experiment established causal correspondence. Likewise,
the proposed roles of token interaction, bilinear coupling, and baseline
anchoring require matched ablations to quantify their individual
contributions.

The inference task differs from those addressed by related computational approaches. Waveform classifiers such as that of Sen et al.
identify occlusion categories directly {[}19{]}, whereas the present
method estimates partial-stenosis parameters by comparing predicted
responses. PINN-based studies have reconstructed unobserved hemodynamic
fields and estimated model parameters by combining observations with
physical constraints {[}13,15,16,17{]}. Here, physical modeling enters
through the reference simulations, and training uses supervised field
and paired-baseline losses. The operator-based reconstruction of
Kaczmarek et al. {[}24{]} provides a closely related methodological
approach; our emphasis is explicit lesion conditioning and candidate
vessel--severity discrimination. These differences in objective and
observation protocol preclude a numerical ranking against published
errors without a common benchmark.

Several limitations constrain translation to patient-specific inference.
The present evidence is based on simulated fields, so the learned
operator inherits the assumptions of the reduced-order reference model.
Its lesion representation describes resistance and lumen changes but
does not resolve three-dimensional jets or local wall stresses. The
dataset also shares 31 anatomical--physiological conditions across
scenario partitions. It therefore tests held-out lesion scenarios within
sampled conditions, rather than unseen patients. A subsequent
anatomy-disjoint evaluation must keep all healthy and diseased fields
from each held-out condition outside training, including the auxiliary
baseline targets. The fixed midpoint and lesion length, single-lesion
assumption, and unchanged network topology impose further restrictions
on the candidate space.

The present inverse problem assumes known conditioning inputs, which would be subject to uncertainty in clinical applications. Errors in radii, pressure inputs,
heart rate, or vessel correspondence could change the ranking of lesion
candidates even when velocity noise is modest. The sensitivity and
calibration studies of Melis, Saxton, and Mansilla Alvarez et al. {[}9,10,11{]} provide relevant approaches for examining which
parameters the observations constrain. Extending the current search to
uncertain inputs will require assessing parameter confounding, not
simply enlarging the candidate set. The present noise experiments
characterize one set of perturbations to simulated velocities; they do
not quantify all sources of acquisition error or model discrepancy. The
more severe noise panels should accordingly be interpreted as
computational stress tests.

The interpretation of the reported performance measures is also subject to statistical limitations. Severity
error conditional on correct localization excludes errors in selecting
the vessel, while an empirical 100\% localization rate in KEY-11 does
not establish perfect clinical sensitivity or specificity. A complete
diagnostic evaluation would require joint location--severity assessment,
explicit healthy-versus-diseased testing, and uncertainty estimates
based on independent conditions and repeated training. The present study
also lacks a complete architectural ablation and a matched comparison
with alternative inverse solvers. Closely competing candidate residuals
may be useful for reporting ambiguous cases, but they require
calibration before being interpreted probabilistically.

Further validation should evaluate entirely held-out anatomical and physiological conditions, uncertain conditioning inputs, and more general lesion configurations, followed by assessment against independent experimental or clinical observations. The clinical-data reconstruction
demonstrated by Sarabian et al. {[}15{]} illustrates a relevant
measurement setting, without constituting validation of the present
model. Within the evaluated setting, the framework provides a controlled computational approach to investigating how a localized narrowing is encoded in distributed
pulsatile signals and which observations distinguish competing lesion
hypotheses.

\section{8. Conclusion}\label{conclusion}

We developed a lesion-conditioned neural operator for estimating cerebral arterial stenosis location and severity from sparse velocity waveforms under known anatomical and physiological conditions.
Localized token injection, cross-vessel interaction, and severity-gated
field decoding linked the lesion parameters to distributed hemodynamic
responses, while paired healthy-field supervision constrained the baseline response.
In the simulated evaluation, localization accuracy depended on measurement placement in addition to the number of observed vessels, and vessel-level analysis
identified errors not apparent from aggregate accuracy. These findings
support further investigation of waveform-based lesion discrimination,
with validation on unseen anatomy, uncertain physiological inputs, and
independent measurements required before patient-specific use.

\section{References}\label{references}

\begin{enumerate}
\def\labelenumi{\arabic{enumi}.}
\tightlist
\item
  Alastruey J, Parker KH, Peiró J, Byrd SM, Sherwin SJ. Modelling the
  circle of Willis to assess the effects of anatomical variations and
  occlusions on cerebral flows. J Biomech. 2007;40(8):1794--1805.
  \url{https://doi.org/10.1016/j.jbiomech.2006.07.008}
\item
  Liu J, Kanoke A, Endo H, Niizuma K, Suito H. Multiscale modeling of
  blood circulation with cerebral autoregulation and network pathway
  analysis for hemodynamic redistribution in the vascular network with
  anatomical variations and stenosis conditions. PLoS Comput Biol.
  2026;22(5):e1013853.
  \url{https://doi.org/10.1371/journal.pcbi.1013853}
\item
  Yuhn C, Oshima M, Chen Y, Hayakawa M, Yamada S. Uncertainty
  quantification in cerebral circulation simulations focusing on the
  collateral flow: Surrogate model approach with machine learning. PLoS
  Comput Biol. 2022;18(7):e1009996.
  \url{https://doi.org/10.1371/journal.pcbi.1009996}
\item
  Reymond P, Merenda F, Perren F, Rüfenacht D, Stergiopulos N.
  Validation of a one-dimensional model of the systemic arterial tree.
  Am J Physiol Heart Circ Physiol. 2009;297(1):H208--H222.
  \url{https://doi.org/10.1152/ajpheart.00037.2009}
\item
  Boileau E, Nithiarasu P, Blanco PJ, Müller LO, Fossan FE, Hellevik LR,
  et al.~A benchmark study of numerical schemes for one-dimensional
  arterial blood flow modelling. Int J Numer Method Biomed Eng.
  2015;31(10):e02732. \url{https://doi.org/10.1002/cnm.2732}
\item
  Hodneland E, Hanson E, Sævareid O, Nævdal G, Lundervold A, Šoltészová
  V, et al.~A new framework for assessing subject-specific whole brain
  circulation and perfusion using MRI-based measurements and a
  multi-scale continuous flow model. PLoS Comput Biol.
  2019;15(6):e1007073.
  \url{https://doi.org/10.1371/journal.pcbi.1007073}
\item
  Ii S, Kitade H, Ishida S, Imai Y, Watanabe Y, Wada S. Multiscale
  modeling of human cerebrovasculature: A hybrid approach using
  image-based geometry and a mathematical algorithm. PLoS Comput Biol.
  2020;16(6):e1007943.
  \url{https://doi.org/10.1371/journal.pcbi.1007943}
\item
  Itu L, Sharma P, Suciu C, Moldoveanu F, Comaniciu D. Personalized
  blood flow computations: A hierarchical parameter estimation framework
  for tuning boundary conditions. Int J Numer Method Biomed Eng.
  2017;33(3):e02803. \url{https://doi.org/10.1002/cnm.2803}
\item
  Melis A, Clayton RH, Marzo A. Bayesian sensitivity analysis of a 1D
  vascular model with Gaussian process emulators. Int J Numer Method
  Biomed Eng. 2017;33(12):e2882. \url{https://doi.org/10.1002/cnm.2882}
\item
  Saxton H, Xu X, Schenkel T, Halliday I. Assessing input parameter
  hyperspace and parameter identifiability in a cardiovascular system
  model via sensitivity analysis. J Comput Sci. 2024;79:102287.
  \url{https://doi.org/10.1016/j.jocs.2024.102287}
\item
  Mansilla Alvarez LA, de Oliveira Mussolin D, Cunha-Lima G, Garzon S,
  Müller LO, Lemos PA, et al.~Towards personalization of 1D blood flow
  models: sensitivity analysis and parameter calibration. Biomech Model
  Mechanobiol. 2026;25:67.
  \url{https://doi.org/10.1007/s10237-026-02094-2}
\item
  Raissi M, Perdikaris P, Karniadakis GE. Physics-informed neural
  networks: A deep learning framework for solving forward and inverse
  problems involving nonlinear partial differential equations. J Comput
  Phys. 2019;378:686--707.
  \url{https://doi.org/10.1016/j.jcp.2018.10.045}
\item
  Kissas G, Yang Y, Hwuang E, Witschey WR, Detre JA, Perdikaris P.
  Machine learning in cardiovascular flows modeling: Predicting arterial
  blood pressure from non-invasive 4D flow MRI data using
  physics-informed neural networks. Comput Methods Appl Mech Eng.
  2020;358:112623. \url{https://doi.org/10.1016/j.cma.2019.112623}
\item
  Arzani A, Wang JX, D'Souza RM. Uncovering near-wall blood flow from
  sparse data with physics-informed neural networks. Phys Fluids.
  2021;33(7):071905. \url{https://doi.org/10.1063/5.0055600}
\item
  Sarabian M, Babaee H, Laksari K. Physics-informed neural networks for
  brain hemodynamic predictions using medical imaging. IEEE Trans Med
  Imaging. 2022;41(9):2285--2303.
  \url{https://doi.org/10.1109/TMI.2022.3161653}
\item
  Sen A, Ghajar-Rahimi E, Aguirre M, Navarro L, Goergen CJ, Avril S.
  Physics-Informed Graph Neural Networks to solve 1-D equations of blood
  flow. Comput Methods Programs Biomed. 2024;257:108427.
  \url{https://doi.org/10.1016/j.cmpb.2024.108427}
\item
  Garay J, Dunstan J, Uribe S, Sahli Costabal F. Physics-informed neural
  networks for parameter estimation in blood flow models. Comput Biol
  Med. 2024;178:108706.
  \url{https://doi.org/10.1016/j.compbiomed.2024.108706}
\item
  Melis A, Moura F, Larrabide I, Janot K, Clayton RH, Narata AP, et
  al.~Improved biomechanical metrics of cerebral vasospasm identified
  via sensitivity analysis of a 1D cerebral circulation model. J
  Biomech. 2019;90:24--32.
  \url{https://doi.org/10.1016/j.jbiomech.2019.04.019}
\item
  Sen A, Navarro L, Avril S, Aguirre M. A data-driven computational
  methodology towards a pre-hospital Acute Ischaemic Stroke screening
  tool using haemodynamics waveforms. Comput Methods Programs Biomed.
  2024;244:107982. \url{https://doi.org/10.1016/j.cmpb.2023.107982}
\item
  Kovachki N, Li Z, Liu B, Azizzadenesheli K, Bhattacharya K, Stuart A,
  et al.~Neural Operator: Learning Maps Between Function Spaces With
  Applications to PDEs. J Mach Learn Res. 2023;24(89):1--97.
  \url{https://www.jmlr.org/papers/v24/21-1524.html}
\item
  Lu L, Jin P, Pang G, Zhang Z, Karniadakis GE. Learning nonlinear
  operators via DeepONet based on the universal approximation theorem of
  operators. Nat Mach Intell. 2021;3:218--229.
  \url{https://doi.org/10.1038/s42256-021-00302-5}
\item
  Li Z, Kovachki N, Azizzadenesheli K, Liu B, Bhattacharya K, Stuart A,
  et al.~Fourier Neural Operator for Parametric Partial Differential
  Equations. International Conference on Learning Representations. 2021.
  \url{https://openreview.net/forum?id=c8P9NQVtmnO}
\item
  Hao Z, Wang Z, Su H, Ying C, Dong Y, Liu S, et al.~GNOT: A General
  Neural Operator Transformer for Operator Learning. Proceedings of the
  40th International Conference on Machine Learning. Proc Mach Learn
  Res. 2023;202:12556--12569.
  \url{https://proceedings.mlr.press/v202/hao23c.html}
\item
  Kaczmarek W, Magdziarz Ibrahim-El-Nur J, Łoś M, Roleder T.
  Reconstructing cerebral hemodynamics from sparse data using Neural
  Operator Transformers. Comput Biol Med. 2025;195:110492.
  \url{https://doi.org/10.1016/j.compbiomed.2025.110492}
\item
  Kennedy McConnell F, Payne S. The dual role of cerebral autoregulation and
  collateral flow in the circle of Willis after major vessel occlusion.
  IEEE Trans Biomed Eng. 2017;64(8):1793--1802.
  \url{https://doi.org/10.1109/TBME.2016.2623710}
\item
  Sherwin SJ, Formaggia L, Peiró J, Franke V. Computational modelling
  of 1D blood flow with variable mechanical properties and its
  application to the simulation of wave propagation in the human
  arterial system. Int J Numer Methods Fluids. 2003;43(6--7):673--700.
  \url{https://doi.org/10.1002/fld.543}
\item
  Filonova V, Arthurs CJ, Vignon-Clementel IE, Figueroa CA.
  Verification of the coupled-momentum method with Womersley's
  Deformable Wall analytical solution. Int J Numer Method Biomed Eng.
  2020;36(2):e3266. \url{https://doi.org/10.1002/cnm.3266}
\item
  Liang F, Fukasaku K, Liu H, Takagi S. A computational model study of
  the influence of the anatomy of the circle of Willis on cerebral
  hyperperfusion following carotid artery surgery. Biomed Eng Online.
  2011;10:84. \url{https://doi.org/10.1186/1475-925X-10-84}
\item
  Young DF, Tsai FY. Flow characteristics in models of arterial
  stenoses---I. Steady flow. J Biomech. 1973;6(4):395--410.
  \url{https://doi.org/10.1016/0021-9290(73)90099-7}
\item
  Young DF, Tsai FY. Flow characteristics in models of arterial
  stenoses---II. Unsteady flow. J Biomech. 1973;6(5):547--559.
  \url{https://doi.org/10.1016/0021-9290(73)90012-2}
\item
  Seeley BD, Young DF. Effect of geometry on pressure losses across
  models of arterial stenoses. J Biomech. 1976;9(7):439--448.
  \url{https://doi.org/10.1016/0021-9290(76)90086-5}
\item
  McKay MD, Beckman RJ, Conover WJ. Comparison of three methods for
  selecting values of input variables in the analysis of output from a
  computer code. Technometrics. 1979;21(2):239--245.
  \url{https://doi.org/10.1080/00401706.1979.10489755}
\item
  Vaswani A, Shazeer N, Parmar N, Uszkoreit J, Jones L, Gomez AN, et
  al.~Attention is all you need. Advances in Neural Information
  Processing Systems. 2017;30.
  \url{https://proceedings.neurips.cc/paper_files/paper/2017/hash/3f5ee243547dee91fbd053c1c4a845aa-Abstract.html}
\item
  Loshchilov I, Hutter F. Decoupled weight decay regularization.
  International Conference on Learning Representations. 2019.
  \url{https://openreview.net/forum?id=Bkg6RiCqY7}
\end{enumerate}

\section{Figures}\label{figures-and-legends}

\includegraphics[width=1\linewidth,height=\textheight,keepaspectratio,alt={Fig 1. Overall computational workflow.}]{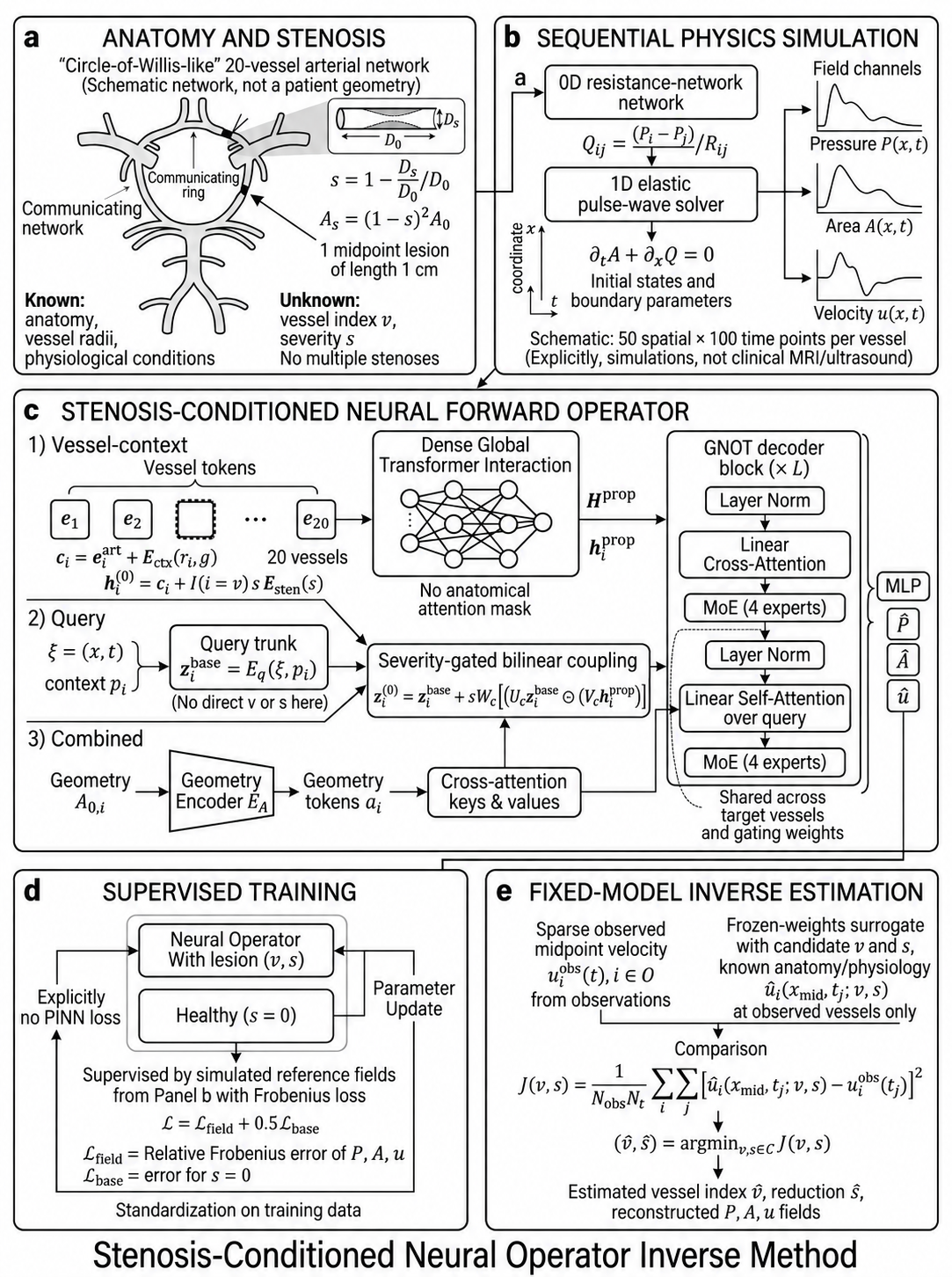}

Figure 1. Computational workflow for stenosis inversion: 0D--1D
simulation generates reference hemodynamic fields; a lesion-conditioned
neural operator learns the forward response; candidate waveform misfit
minimization estimates stenosis location and severity from sparse
velocity observations.

\includegraphics[width=1\linewidth,height=\textheight,keepaspectratio,alt={Fig 2. Representative forward waveform comparison.}]{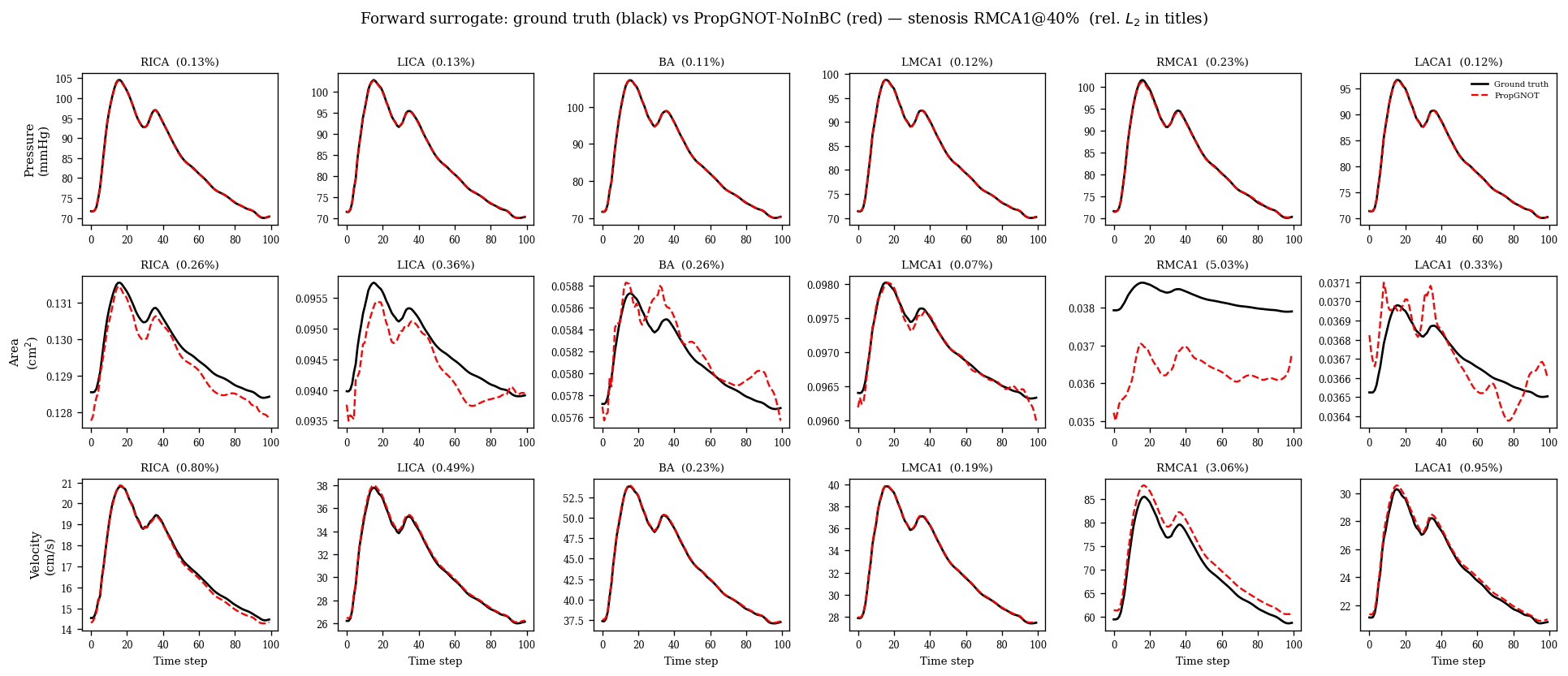}

Figure 2. Reference and predicted pressure, velocity, and area waveforms
at representative vessels for an RMCA1 stenosis with 40\% diameter
reduction. Predictions use the prescribed lesion parameters;
panel annotations report relative waveform errors.

\includegraphics[width=1\linewidth,height=\textheight,keepaspectratio,alt={Vessel-level forward error distribution.}]{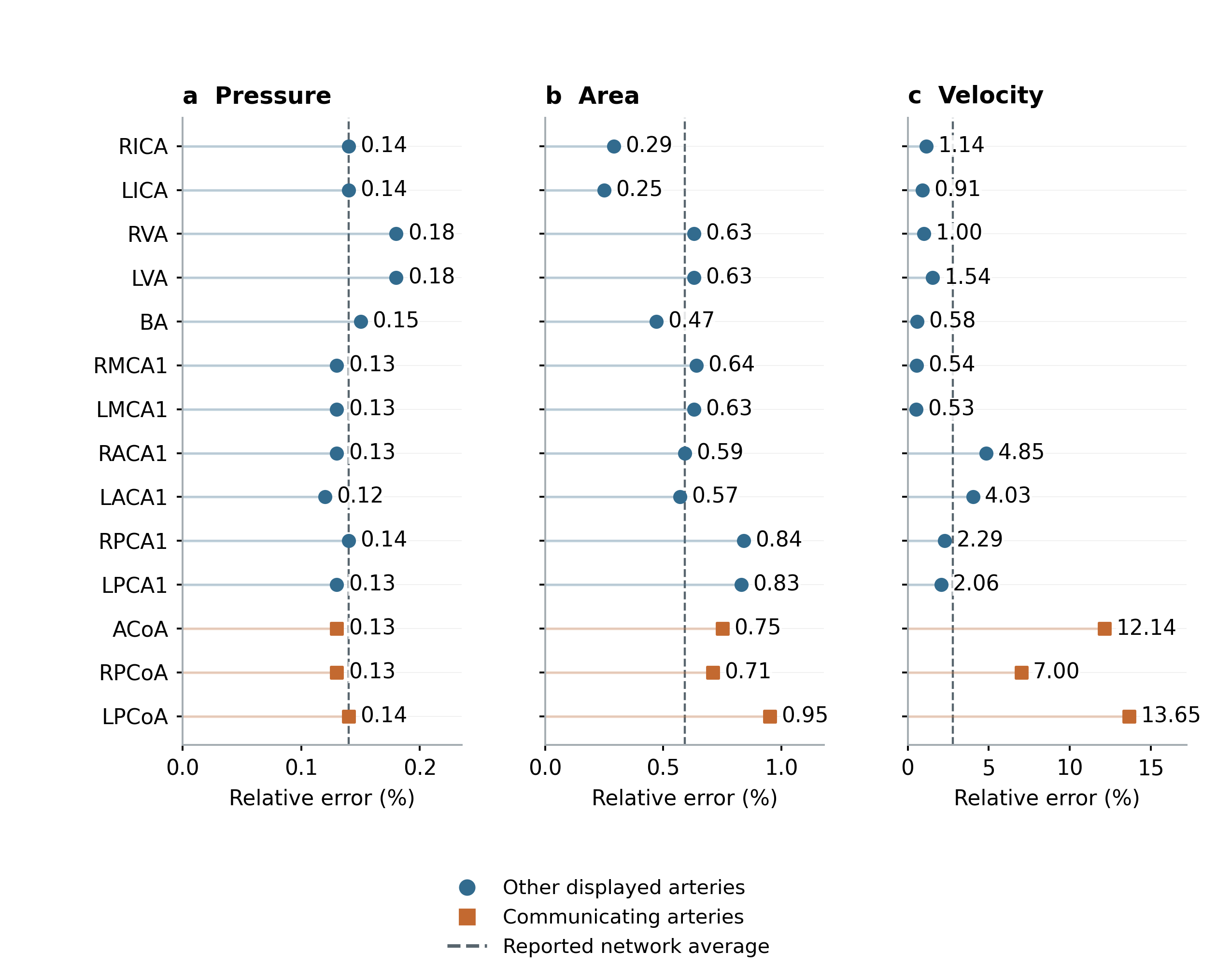}

Figure 3. Relative reconstruction errors for the 14 arteries in Table 2:
(a) pressure; (b) area; (c) velocity. Orange squares denote communicating
arteries and blue circles denote the remaining arteries. Dashed lines
indicate the reported full-network averages.

\includegraphics[width=0.65\linewidth,height=\textheight,keepaspectratio,alt={Fig 4. Severity estimation under 8v2 without added noise.}]{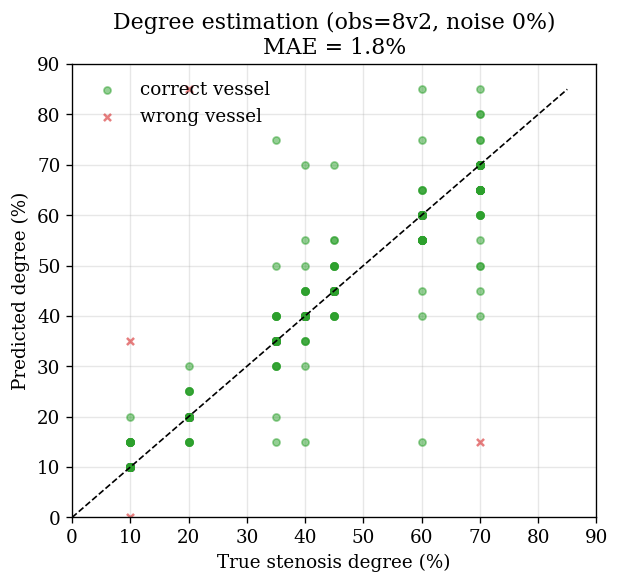}

Figure 4. Estimated versus reference diameter reduction under the 8v2
configuration without added noise. The diagonal denotes exact agreement;
markers distinguish correct and incorrect vessel localization.

\includegraphics[width=1\linewidth,height=\textheight,keepaspectratio,alt={Fig 5A. Localization versus observation noise.}]{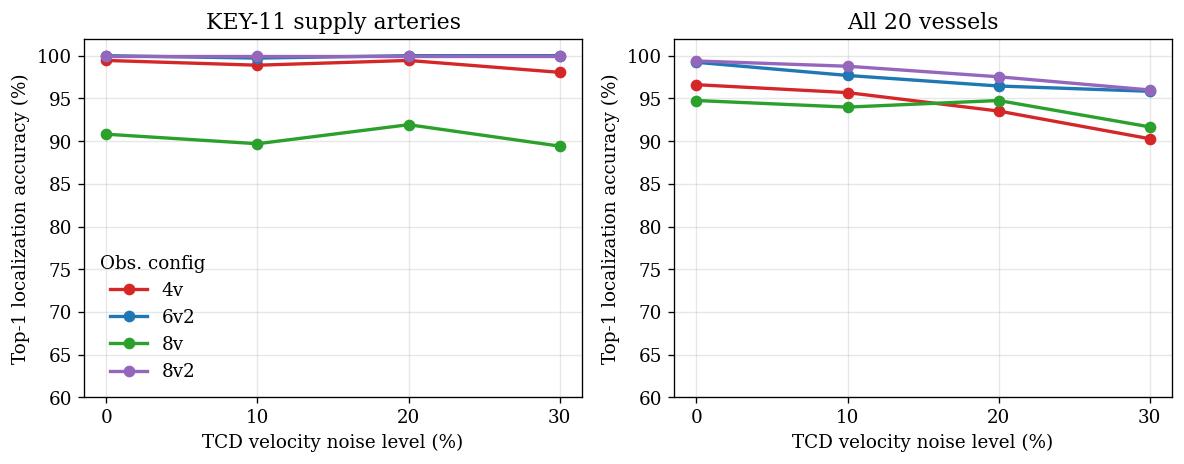}

\includegraphics[width=0.7\linewidth,height=\textheight,keepaspectratio,alt={Fig 5B. Severity error versus observation noise.}]{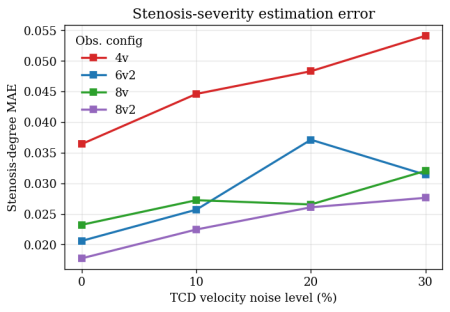}

Figure 5. Inverse estimation across observation configurations and noise
levels: (A) Top-1 localization accuracy for KEY-11 and all 20 candidate
vessels; (B) mean absolute error in stenosis severity.

\includegraphics[width=1\linewidth,height=\textheight,keepaspectratio,alt={Fig 6. Localization stratified by severity and noise.}]{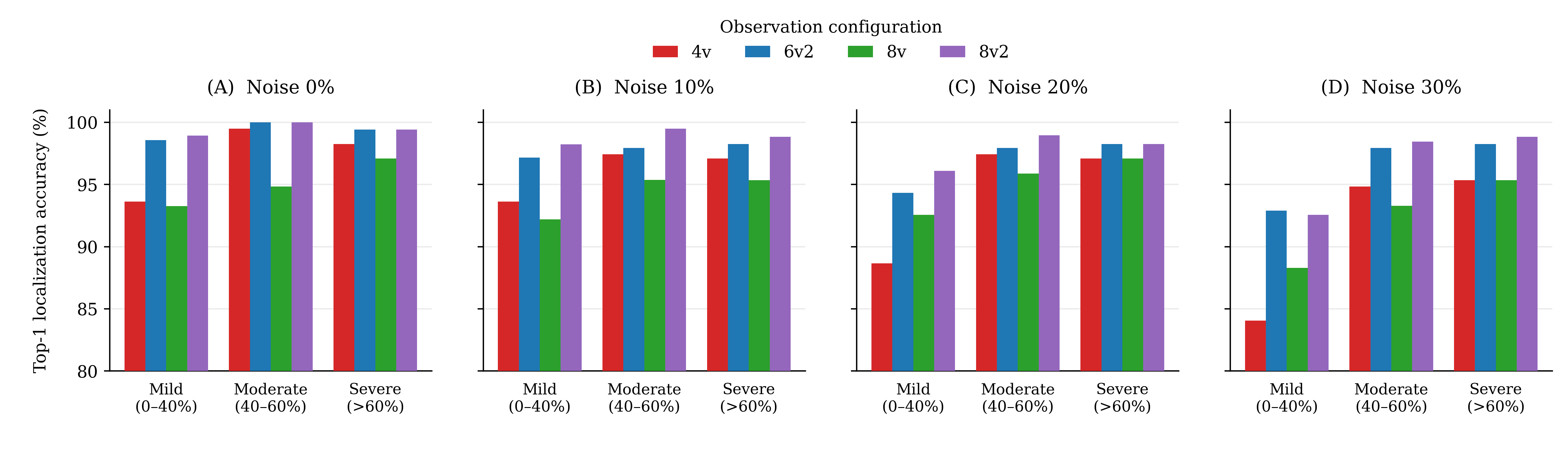}

Figure 6. Top-1 localization accuracy stratified by stenosis severity,
observation configuration, and noise level. Severity groups correspond
to the diameter-reduction intervals indicated in each panel. Vertical axes
start at 80\%.

\includegraphics[width=0.7\linewidth,height=\textheight,keepaspectratio,alt={Fig 7A. Per-vessel mislocalization across noise settings.}]{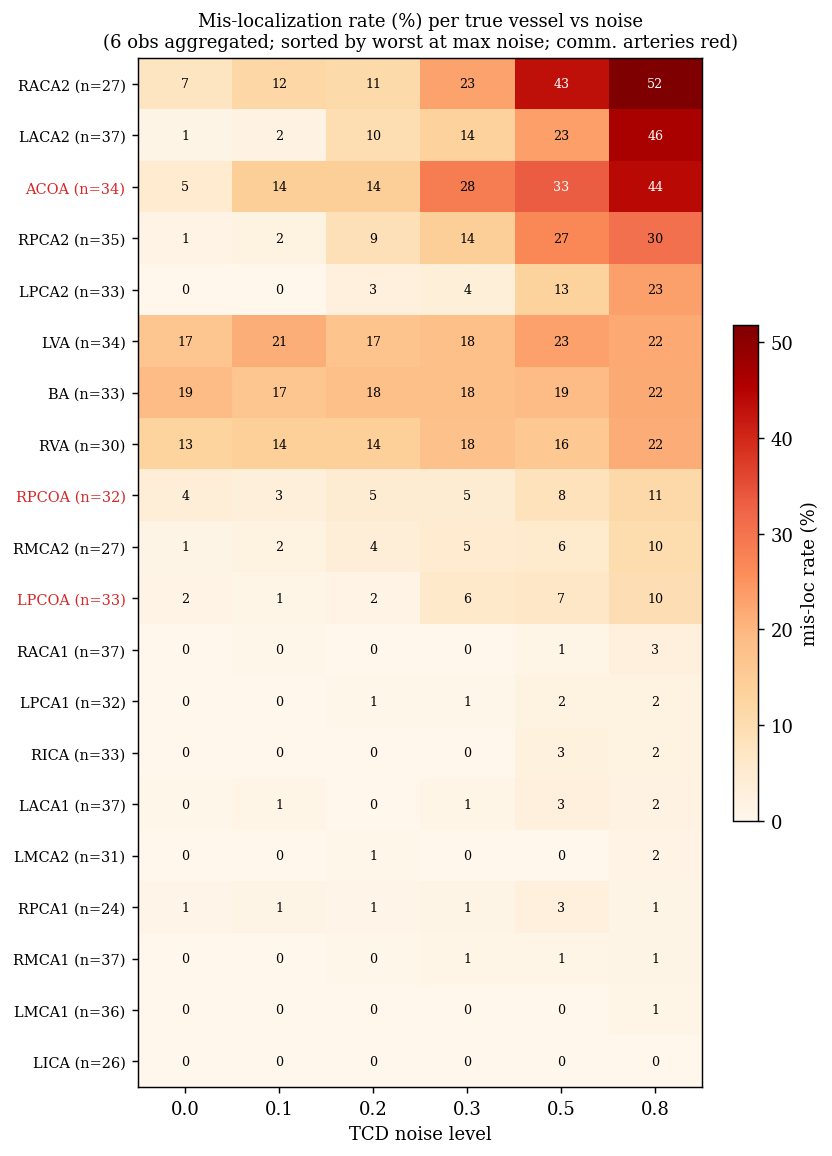}

\includegraphics[width=0.8\linewidth,height=\textheight,keepaspectratio,alt={Fig 7B. Residual separation and mislocalization at the 30\% setting.}]{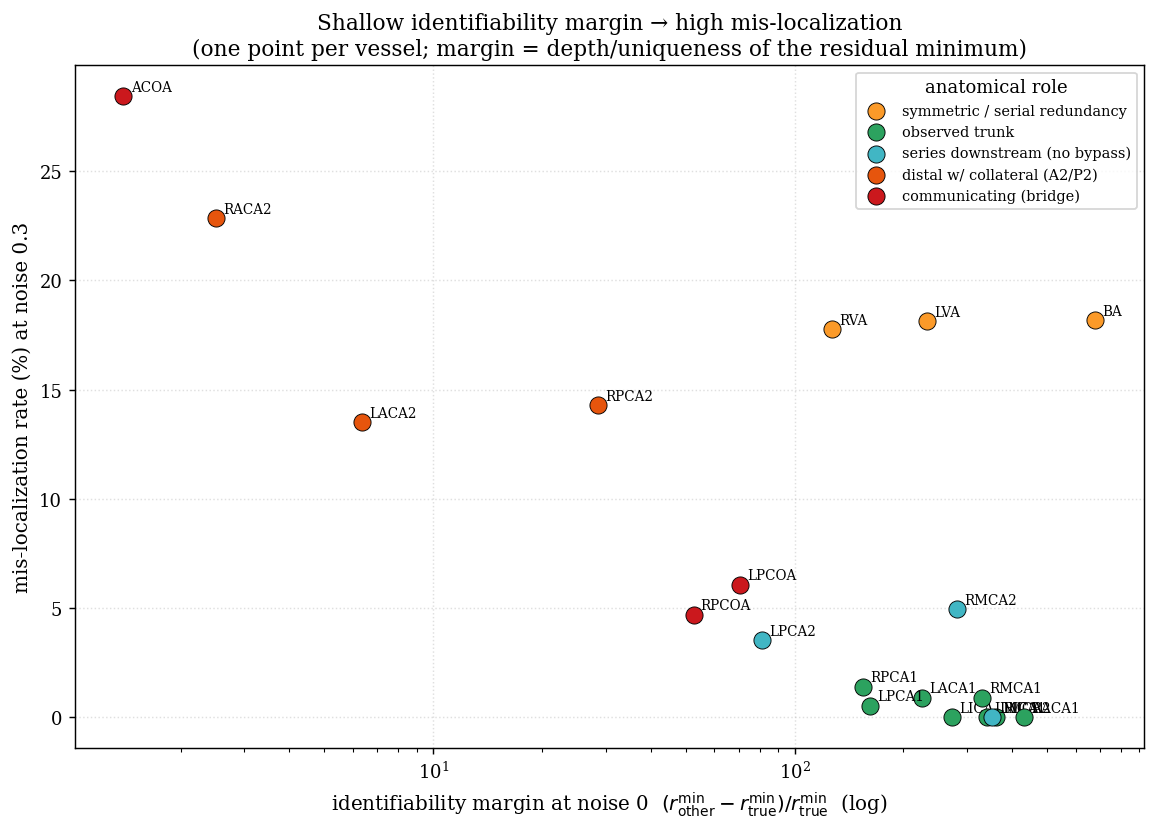}

\includegraphics[width=0.8\linewidth,height=\textheight,keepaspectratio,alt={Fig 7C. Residual separation and mislocalization at the 80\% setting.}]{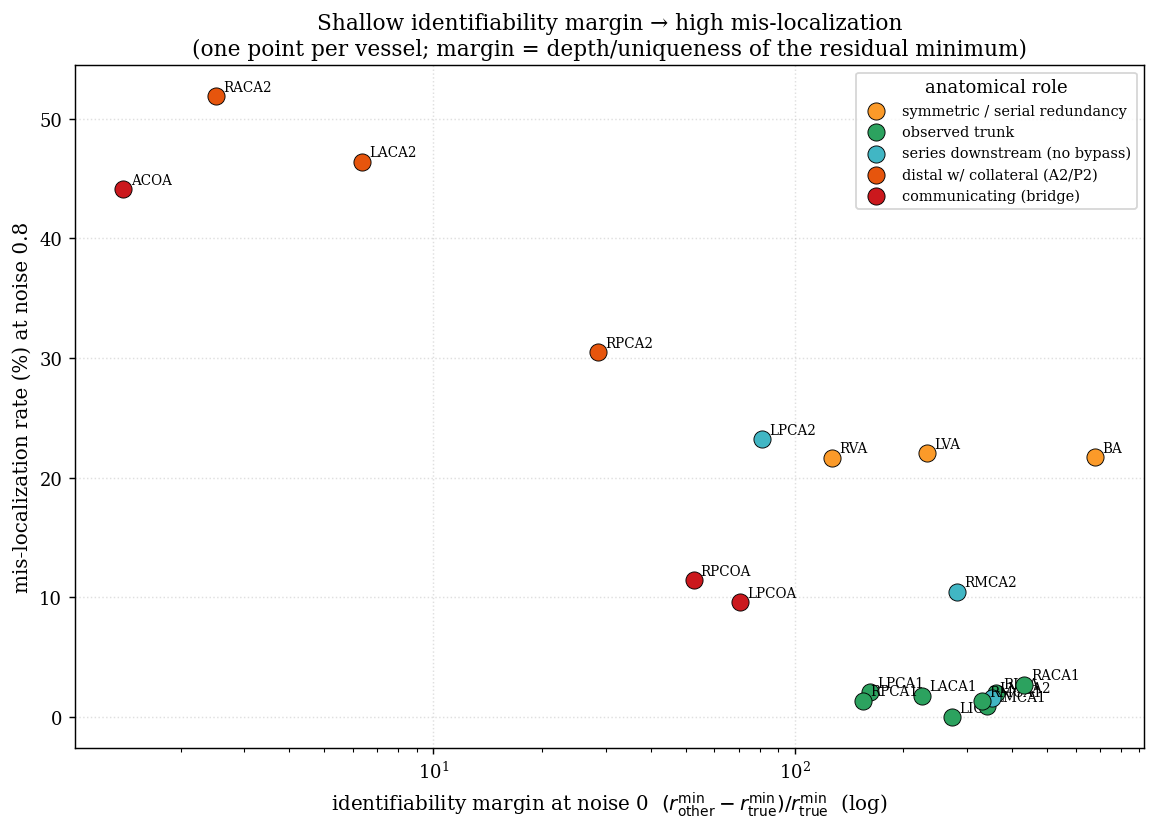}

Figure 7. Vessel-specific localization errors and candidate residual
separation: (A) mislocalization rates across noise levels, pooled over
six observation configurations; (B,C) noiseless residual separation
versus mislocalization at 30\% and 80\% noise, respectively.

\includegraphics[width=0.85\linewidth,height=\textheight,keepaspectratio,alt={Fig 8. Learned vessel-token attention.}]{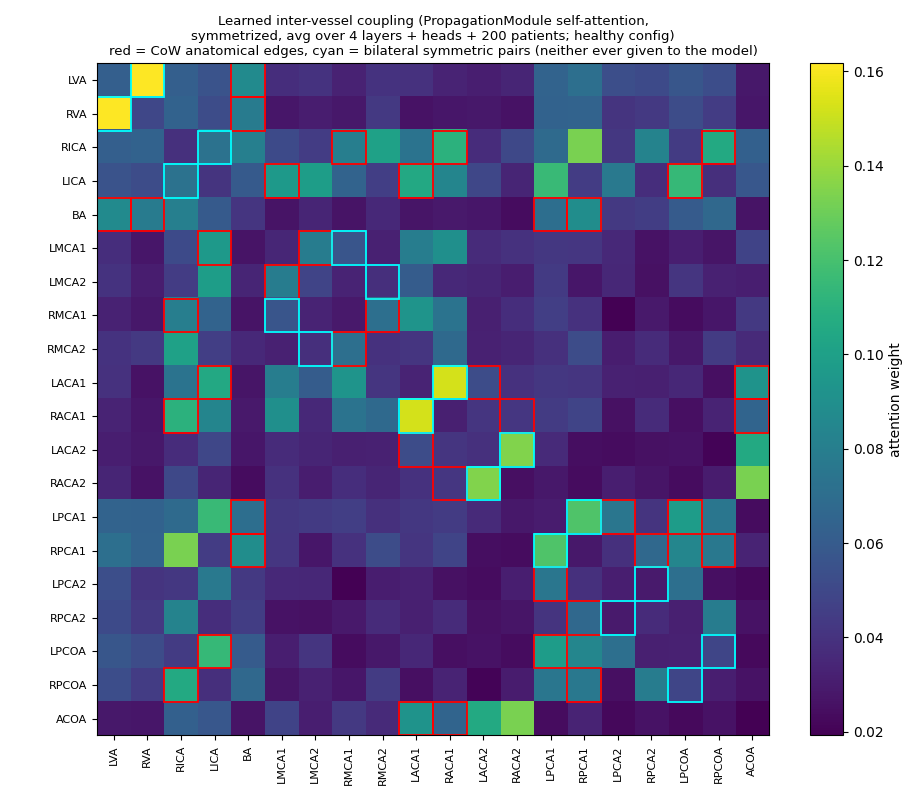}

Figure 8. Aggregated, symmetrized self-attention among the 20 vessel
tokens in the propagation module. Overlaid boxes identify anatomical
connections and bilateral vessel pairs.

\section{Supporting information}\label{supporting-information}

\subsection{S1 Appendix. Model and evaluation
settings}\label{s1-appendix.-confirmed-model-and-evaluation-settings}

{\def\LTcaptype{none} 
\begin{longtable}[]{@{}
  >{\raggedright\arraybackslash}p{(\linewidth - 2\tabcolsep) * \real{0.5000}}
  >{\raggedright\arraybackslash}p{(\linewidth - 2\tabcolsep) * \real{0.5000}}@{}}
\toprule\noalign{}
\begin{minipage}[b]{\linewidth}\raggedright
Component
\end{minipage} & \begin{minipage}[b]{\linewidth}\raggedright
Setting stated in the study
\end{minipage} \\
\midrule\noalign{}
\endhead
\bottomrule\noalign{}
\endlastfoot
Target vessels & 20 \\
Lesion condition & One vessel index and diameter reduction \\
Lesion geometry & Midpoint, 1 cm length \\
Anatomy--physiology conditions & 31 \\
Scenario partition & Approximately 70\% / 15\% / 15\% \\
Output grid & 50 spatial points × 100 time points per vessel \\
Predicted channels & Pressure, area, velocity \\
Propagation & Global Transformer interaction among vessel tokens \\
Anatomical adjacency mask & None in the neural propagation module \\
Geometry input & Reference cross-sectional area function \\
Coupling & Severity-gated bilinear interaction \\
Decoder & Shared GNOT blocks with geometric mixtures of experts \\
Experts & Four per expert group \\
Standardization & Training statistics; continuous channels \\
Primary training loss & Relative Frobenius field error \\
Auxiliary training loss & Paired healthy-baseline relative Frobenius
error \\
Baseline weight & 0.5 \\
Optimizer & AdamW \\
Learning-rate schedule & Cosine \\
Epochs & 200 \\
Batch size & 1 \\
Training hardware & One NVIDIA H100 GPU \\
Inverse observations & Midpoint velocity waveforms over one cardiac
cycle \\
Inverse residual & Mean squared velocity discrepancy \\
Inverse conditioning & Known anatomy and physiology \\
\end{longtable}
}

\subsection{S1 Fig. Training and candidate-search
algorithms}\label{s1-fig.-training-and-candidate-search-algorithms}

\includegraphics[width=1\linewidth,height=\textheight,keepaspectratio,alt={Training and candidate-search algorithms.}]{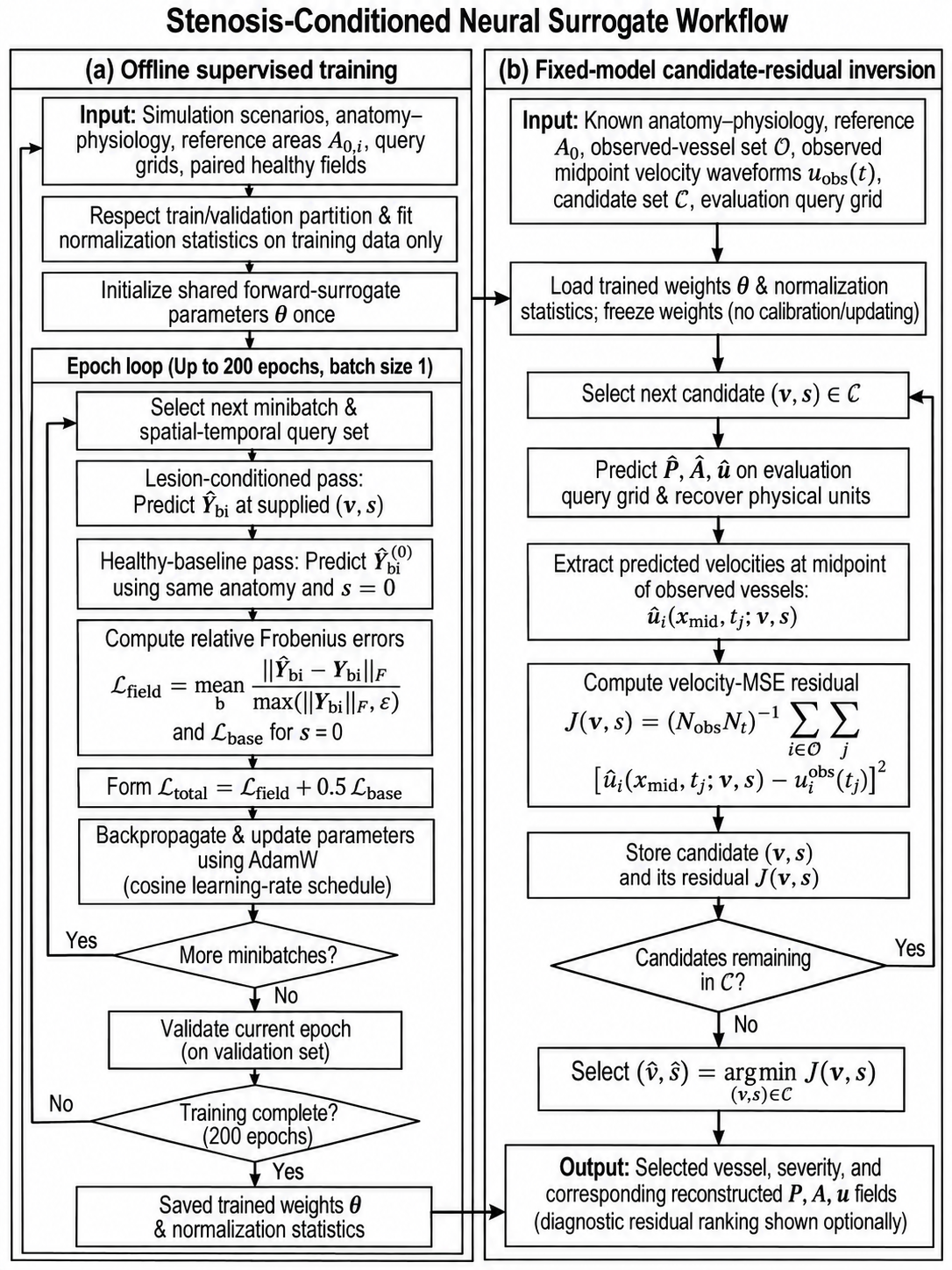}

Figure S1. Flowchart of the computational control flow: (a) offline training optimizes parameters; (b) frozen-model validation determines candidate parameters via waveform misfit minimization.

\end{document}